\documentclass[preprint2]{aastex631}

\usepackage{verbatim}
\usepackage{multirow}
\usepackage{booktabs}
\usepackage{amsmath}
\shorttitle{Reprocessing Models for AGN}
\shortauthors{Akiba et al.}

\graphicspath{{./}{Figures/}}

\def\tanote#1{{\color{black}#1}}

\begin{document}

\title{Reprocessing Models for the Optical Light Curves of Hypervariable Quasars from the Sloan Digital Sky Survey Reverberation Mapping Project}

\author[0000-0002-0647-718X]{Tatsuya Akiba}
\affiliation{JILA, University of Colorado and National Institute of Standards and Technology, 440 UCB, Boulder, CO 80309-0440, USA}
\affiliation{Department of Astrophysical and Planetary Sciences, University of Colorado, 391 UCB, Boulder, CO 80309, USA}
\email{tatsuya.akiba@colorado.edu}

\author[0000-0003-3903-0373]{Jason Dexter}
\affiliation{JILA, University of Colorado and National Institute of Standards and Technology, 440 UCB, Boulder, CO 80309-0440, USA}
\affiliation{Department of Astrophysical and Planetary Sciences, University of Colorado, 391 UCB, Boulder, CO 80309, USA}

\author[0000-0002-0167-2453]{William Brandt}
\affiliation{Department of Astronomy and Astrophysics, 525 Davey Lab, The Pennsylvania State University, University Park, PA 16802, USA}
\affiliation{Institute for Gravitation and the Cosmos, The Pennsylvania State University, University Park, PA 16802, USA}
\affiliation{Department of Physics, 104 Davey Laboratory, The Pennsylvania State University, University Park, PA 16802, USA}

\author[0000-0001-6947-5846]{Luis C. Ho}
\affiliation{Kavli Institute for Astronomy and Astrophysics, Peking University, Beijing 100871, China}
\affiliation{Department of Astronomy, School of Physics, Peking University, Beijing 100871, China}

\author[0000-0002-0957-7151]{Yasaman Homayouni}
\affiliation{Department of Astronomy and Astrophysics, 525 Davey Lab, The Pennsylvania State University, University Park, PA 16802, USA}
\affiliation{Space Telescope Science Institute, 3700 San Martin Drive, Baltimore, MD 21218, USA}

\author{Donald P. Schneider}
\affiliation{Department of Astronomy and Astrophysics, The Pennsylvania State University,
   University Park, PA 16802}
\affiliation{Institute for Gravitation and the Cosmos, The Pennsylvania State University,
   University Park, PA 16802}

\author[0000-0001-8610-5732]{Yue Shen}
\affiliation{Department of Astronomy, University of Illinois at Urbana-Champaign, Urbana, IL, 61801, USA}
\affiliation{National Center for Supercomputing Applications, University of Illinois at Urbana-Champaign, Urbana, IL, 61801, USA}

\author[0000-0002-1410-0470]{Jonathan R. Trump}
\affil{Department of Physics, 196A Auditorium Road, Unit 3046, University of Connecticut, Storrs, CT 06269, USA}

\begin{abstract}

We explore reprocessing models for a sample of 17 hypervariable quasars, taken from the Sloan Digital Sky Survey Reverberation Mapping (SDSS-RM) project, which all show coordinated optical luminosity hypervariability with amplitudes of factors $\gtrsim 2$ between 2014 and 2020. We develop and apply reprocessing models for quasar light curves in simple geometries that are likely to be representative of quasar inner environments. In addition to the commonly investigated thin-disk model, we include the thick-disk and hemisphere geometries. The thick-disk geometry could, for instance, represent a magnetically-elevated disk, whereas the hemisphere model can be interpreted as a first-order approximation for any optically-thick out-of-plane material caused by outflows/winds, warped/tilted disks, etc. Of the 17 quasars in our sample, eleven are best-fit by a hemisphere geometry, five are classified as thick disks, and both models fail for just one object. We highlight the successes and shortcomings of our thermal reprocessing models in case studies of four quasars that are representative of the sample. While reprocessing is unlikely to explain all of the variability we observe in quasars, we present our classification scheme as a starting point for revealing the likely geometries of reprocessing for quasars in our sample and hypervariable quasars in general.

\end{abstract}


\section{Introduction} \label{sec:intro}

According to standard thin accretion disk theory \citep{shakura1973,pringle1981}, the expected timescale for quasars to change their intrinsic accretion power is of order the inflow timescale, $t_{\rm inflow} \sim 10^4 $--$ 10^6$ years. Observationally, the continuum UV/optical emission of quasars shows fluctuations on much shorter timescales of months to years \citep[e.g.,][]{kelly2009,macleod2010}. This variability is typically on the order $10 $--$ 20 \%$ in root-mean-square (rms) and is correlated over a wide range of UV/optical wavelength bands \citep[e.g.,][]{clavel1991,krolik1991}, following a general pattern of being bluer when brighter \citep[e.g.,][]{Wilhite05, macleod2016}. These luminosity variations are usually attributed to thermal reprocessing: emission from near the black hole being absorbed and re-emitted by the surrounding accretion disk \citep[e.g.,][]{Wanders97, Collier99, Fausnaugh16, Edelson15, McHardy18}. Instrinsic variations in the incident ionizing continuum flux translate into correlated changes in the reprocessed emission from the disk and the broad line region, and these interband variability signatures also allow for disk-size measurements \citep[e.g.,][]{Fausnaugh16}. Thermal reprocessing can explain small variations reasonably well --- we can think of the continuum emission of quasars as ``flickering'' small-amplitude variability due to reprocessing sitting on top of intrinsic quasar variability operating on a longer timescale.

However, the Sloan Digital Sky Survey \citep[SDSS; ][]{sdss}, amongst other observational endeavors, has recently uncovered a surprising number of so-called ``hypervariable'' quasars \citep{macleod2016,rumbaugh2018}. These objects exhibit optical luminosity variations of factors $\gtrsim 2$ on timescales of months to years \citep{rumbaugh2018}. These large-amplitude, rapid changes in luminosity have long been studied in nearby Seyferts, but are now being observed in much more luminous quasars. These large variations do not fit in with our previous physical picture. Significant changes in intrinsic accretion power are predicted to occur on much longer timescales, and reprocessing in a thin disk should be of much smaller amplitudes due to geometric dilution: less of the intrinsic variability is effectively captured and re-emitted at redder wavelength bands as one moves progressively farther out in the disk.

Several explanations for this phenomenon have been proposed: 1. Quasars could be subject to disk instabilities with rapid state transitions \citep[e.g.,][]{noda2018}; 2. The inflow time through the disk could be much faster than commonly assumed \citep[e.g.,][]{dexterbegelman2019}; or 3. The entire optical spectra could be powered by reprocessing \citep[e.g.,][]{shappee2014,lawrence2018}. The last theory is particularly attractive as quasar variability is usually correlated with ``blue'' leading ``red'', i.e. shorter wavelength bands exhibiting changes at earlier times \cite[e.g.,][]{clavel1991,peterson1991}. Hypervariable quasars have also been found to exhibit the same behavior at much larger variability amplitude \citep[e.g.,][]{dexter2019}. In the reprocessing picture, this property is explained by light travel time-delays between the central source and the disk: the inner parts of the disk reflect the variability earlier than the outer parts. In fact, quasar variability has been successfully modeled by this mechanism for particular cases \citep[e.g.,][]{shappee2014}.

Nonetheless, hypervariability remains challenging to explain. In a thin-disk geometry, reprocessing alone has difficulty generating large fluctuations across a wide range of wavelengths \citep{dexter2019}. Furthermore, there are several arguments against reprocessing as an explanation for hypervariability in quasars. For some local AGNs with extensive simultaneous X-ray and UV/optical monitoring, the X-ray variability is insufficient to drive the large amplitude UV/optical variability \citep[e.g.,][]{McHardy2014}. In addition, the X-ray/UV correlations are found to be much weaker than those in the UV/optical \citep[e.g.,][]{edelson2015, buisson2018, Edelson19}. If these results also hold for the SDSS-RM quasars, reprocessing is unlikely to account for all of the hypervariability we see in the UV/optical. Another argument against reprocessing as the origin of UV/optical variability is related to variability timescales. The damping timescale of UV/optical light curves from a damped random walk model is consistent with the thermal timescale of the disk, which suggests that the long-term UV/optical variability is likely from thermal fluctuations of the disk rather than reprocessing \citep[e.g.,][]{kelly2009}.

In this paper, we explore the potential for thermal reprocessing models to explain the optical spectra of a sample of SDSS hypervariable quasars by considering alternate geometries with large covering areas. We examine the hemisphere geometry, where we imagine a reprocessing hemisphere at a given radius from the central source, and a thick disk, i.e. an accretion disk with a substantial height-to-radius ($H/R$) ratio. In the latter, we use the ``lamp post'' model \citep[e.g.,][]{cackett2007, shappee2014}, where the central source of intrinsic variability is slightly elevated with respect to the midplane of the disk. In both cases, we effectively reduce geometric dilution such that most, if not all, of the emission from near the black hole can be thermally reprocessed.

In Section \ref{sec:observations}, we describe the selection of our hypervariable quasar sample. In Section \ref{sec:methods}, we explain our data analysis and characterize the quantitative models associated with our two geometries. The model fits and classification results are described in Section \ref{sec:results}, and our conclusions are presented in Section \ref{sec:conclusion} with a discussion following in Section \ref{sec:discussion}.

% \begin{deluxetable*}
% \begin{tabular}{ ||c||c|c|c|c|c|| }
%  \toprule
%  \hline
% RM-ID & RA (J2000) & Dec (J2000) & $z$ & log $M_{\rm BH}/M_{\odot}$ & PSF mag (u) \\
% \hline 
% 12 & 213.4822 & 53.2006 & 1.58 & 8.30 & 22.35 \\
% 17 & 213.3511 & 53.0908 & 0.46 & 8.4 & 21.15 \\
% 32 & 213.3064 & 52.9306 & 1.71 & 7.60 & 20.53 \\
% 105 & 214.5063 & 52.8669 & 1.16 & 9.05 & 20.17 \\
% 112 & 212.8857 & 52.8532 & 1.40 & 9.23 & 19.96 \\
% 143 & 213.6296 & 53.7088 & 1.23 & 8.77 & 20.87 \\
% 160 & 212.6719 & 53.3136 & 0.36 & 8.2 & 19.73 \\
% 194 & 213.1297 & 52.4422 & 1.56 & 8.98 & 21.90 \\
% 303 & 214.6259 & 52.3701 & 0.82 & 8.3 & 21.14 \\
% 309 & 214.8457 & 53.6932 & 1.32 & 8.76 & 21.25 \\
% 346 & 214.6820 & 53.8607 & 1.59 & 8.68 & 21.78 \\
% 434 & 212.2986 & 52.3973 & 1.55 & 8.69 & 20.89 \\
% 559 & 215.7504 & 53.2819 & 1.22 & 8.43 & 21.35 \\
% 597 & 215.2584 & 52.1978 & 1.20 & 8.46 & 21.53 \\
% 714 & 215.9572 & 52.6510 & 0.92 & 8.9 & 20.40 \\
% 768 & 212.3154 & 53.4561 & 0.26 & 8.7 & 20.43 \\
% 839 & 213.4954 & 54.4517 & 0.98 & 9.1 & 21.33 \\
%  \hline
%  \bottomrule
% \end{tabular}
% \caption{Summary of SDSS-RM objects analyzed in this study. \tanote{The redshift ($z$) is the improved systemic redshift and the magnitude is the point spread function (PSF) magnitude in the SDSS u-band taken from \citet{shen2019}.} Black hole masses are calculated from H$\beta$ \citep{grier2017}, C IV \citep{grier2019}, or Mg II linewdiths \citep{shen2011, homayouni2020} in this order of preference where available.}
% \label{table:rmid_sample}
% \end{deluxetable*}

\begin{deluxetable*}{cccccc}
\tablehead{\colhead{RM-ID} & \colhead{RA (J2000)} & \colhead{Dec (J2000)} & \colhead{$z$} & \colhead{log $M_{\rm BH}/M_{\odot}$} & \colhead{PSF mag (u)}} 

\startdata
12 &  213.4822 &  53.2006 &  1.58 &  8.30 &  22.35 \\
17 &  213.3511 &  53.0908 &  0.46 &  8.4 &  21.15 \\
32 &  213.3064 &  52.9306 &  1.71 &  7.60 &  20.53 \\
105 &  214.5063 &  52.8669 &  1.16 &  9.05 &  20.17 \\
112 &  212.8857 &  52.8532 &  1.40 &  9.23 &  19.96 \\
143 &  213.6296 &  53.7088 &  1.23 &  8.77 &  20.87 \\
160 &  212.6719 &  53.3136 &  0.36 &  8.2 &  19.73 \\
194 &  213.1297 &  52.4422 &  1.56 &  8.98 &  21.90 \\
303 &  214.6259 &  52.3701 &  0.82 &  8.3 &  21.14 \\
309 &  214.8457 &  53.6932 &  1.32 &  8.76 &  21.25 \\
346 &  214.6820 &  53.8607 &  1.59 &  8.68 &  21.78 \\
434 &  212.2986 &  52.3973 &  1.55 &  8.69 &  20.89 \\
559 &  215.7504 &  53.2819 &  1.22 &  8.43 &  21.35 \\
597 &  215.2584 &  52.1978 &  1.20 &  8.46 &  21.53 \\
714 &  215.9572 &  52.6510 &  0.92 &  8.9 &  20.40 \\
768 &  212.3154 &  53.4561 &  0.26 &  8.7 &  20.43 \\
839 &  213.4954 &  54.4517 &  0.98 &  9.1 &  21.33 \\
\enddata

\caption{Summary of SDSS-RM objects analyzed in this study. \tanote{The redshift ($z$) is the improved systemic redshift and the magnitude is the point spread function (PSF) magnitude in the SDSS u-band taken from \citet{shen2019}.} Black hole masses are calculated from H$\beta$ \citep{grier2017}, C IV \citep{grier2019}, or Mg II linewidths \citep{shen2011, homayouni2020} in this order of preference where available.}
\label{table:rmid_sample}
\end{deluxetable*}

\section{Observations} \label{sec:observations}

The 17 hypervariable quasars in our sample were all targets of the SDSS Reverberation Mapping (SDSS-RM) campaign \citep{shen2015} , which began with SDSS-III \citep{eisenstein11} and continued as part of SDSS-IV \citep{blanton2017}. SDSS-RM has dramatically expanded the quasar parameter space in terms of spectroscopic variability, accretion rate, redshift, and multiwavelength properties of quasars due to its simpler (solely magnitude-limited) selection criteria \citep{shen2015}. The SDSS-RM program observed 849 quasars in a 7 deg$^2$ field which corresponds to that of the Pan-STARRS1 Medium Deep Field MD07 \citep[][]{tonry2012}.  The monitoring includes spectroscopy with the Apache Point 2.5-meter SDSS telescope \citep{gunn2006} and photometry from the CFHT and Bok telescopes. Pan-STARRS 1 photometry is also available from 2010--2013, along with GALEX observations and two XMM-Newton observations in 2017.

Our sample consists of the SDSS-RM quasars that displayed $> 1$ mag $g$-band variability in the Pan-STARRS observations (88 total). The sample was further refined to select objects showing $\gtrsim 2$ variability during SDSS-RM, which ends up as a selection on sufficiently high data quality and rms variability. The resulting subset of 17 objects' designations and J2000 coordinates are given in Table \ref{table:rmid_sample}. Of the objects in our sample, only RM 17 is radio-loud \citep{shen2019}. The analysis presented here uses only SDSS-IV BOSS spectra. See \citet{shen2019} for more details on the SDSS-RM sample and its properties.

The SDSS-RM spectra analyzed here were obtained with the BOSS spectrographs \citep{smee2013} between January 2014 and February 2020. We use data from 85 epochs, with a median cadence of only 4 days in 2014 and 16 days in the other years over 7 months of observing per year. The exposure time was typically 2 hr, and the data were first processed by the BOSS pipeline, followed by a custom scheme to improve spectrophotometry and sky subtraction \citep[for technical  details on the SDSS-RM spectroscopy, see][]{shen2015}. The typical absolute spectrophotometric accuracy achieved is $\sim 5\%$ \citep{shen2015}. More details on the spectroscopic data and analysis can be found in \citet{shen2018} and \citet{grier2019}. From the spectra, we estimate continuum luminosities as mean values over narrow bands of width $\approx 50$ \AA, avoiding emission lines in the time-averaged spectrum. All wavelength bands considered in this study are presented in Table \ref{table:wavelength_bands}. We have not attempted to remove weak, narrow emission lines or the Fe II pseudo-continuum. 

\begin{table*}[t!]
\begin{center}
\begin{tabular}{ ||c||cccccccccc|| } 
 \hline
RM-ID & \multicolumn{10}{c||}{Rest-Frame Wavelength Bands (\AA)} \\
\hline
12 & \textit{1455.0} & \textbf{1750.0} & 2175.0 & 2687.5 & 2975.0 & 3350.0 & \textit{3550.0} & \textit{4025.0} & {} & {} \\
17 & \textit{2687.5} & \textbf{2975.0} & 3350.0 & 3550.0 & 4025.0 & 4500.0 & 5150.0 & 5550.0 & 6050.0 & \textit{6900.0} \\
32 & \textit{1322.5} & \textit{1455.0} & \textbf{1750.0} & 2175.0 & 2687.5 & 2975.0 & \textit{3350.0} & \textit{3550.0} & {} & {} \\
105 & \textit{1750.0} & \textbf{2175.0} & 2687.5 & 2975.0 & 3350.0 & 3550.0 & 4025.0 & \textit{4500.0} & {} & {} \\
112 & \textbf{1750.0} & 2175.0 & 2687.5 & 2975.0 & 3350.0 & 3550.0 & \textit{4025.0} & {} & {} & {} \\
143 & \textit{1750.0} & \textbf{2175.0} & 2687.5 & 2975.0 & 3350.0 & 3550.0 & 4025.0 & \textit{4500.0} & {} & {} \\
160 & \textit{2687.5} & \textit{2975.0} & \textbf{3350.0} & 3550.0 & 4025.0 & 4500.0 & 5150.0 & 5550.0 & 6050.0 & \textit{6900.0} \\
194 & \textit{1455.0} & \textbf{1750.0} & 2175.0 & 2687.5 & 2975.0 & 3350.0 & \textit{3550.0} & \textit{4025.0} & {} & {} \\
303 & \textit{2175.0} & \textbf{2687.5} & 2975.0 & 3350.0 & 3550.0 & 4025.0 & 4500.0 & \textit{5150.0} & \textit{5550.0} & {} \\
309 & \textit{1750.0} & \textbf{2175.0} & 2687.5 & 2975.0 & 3350.0 & 3550.0 & \textit{4025.0} & \textit{4500.0} & {} & {} \\
346 & \textit{1455.0} & \textbf{1750.0} & 2175.0 & 2687.5 & 2975.0 & 3350.0 & \textit{3550.0} & \textit{4025.0} & {} & {} \\
434 & \textit{1455.0} & \textbf{1750.0} & 2175.0 & 2687.5 & 2975.0 & 3350.0 & \textit{3550.0} & \textit{4025.0} & {} & {} \\
559 & \textit{1750.0} & \textbf{2175.0} & 2687.5 & 2975.0 & 3350.0 & 3550.0 & 4025.0 & \textit{4500.0} & {} & {} \\
597 & \textit{1750.0} & \textbf{2175.0} & 2687.5 & 2975.0 & 3350.0 & 3550.0 & 4025.0 & \textit{4500.0} & {} & {} \\
714 & \textbf{2175.0} & 2687.5 & 2975.0 & 3350.0 & 3550.0 & 4025.0 & 4500.0 & \textit{5150.0} & {} & {} \\
768 & \textit{2975.0} & \textbf{3350.0} & 3550.0 & 4025.0 & 4500.0 & 5150.0 & 5550.0 & 6050.0 & 6900.0 & {} \\
839 & \textbf{2175.0} & 2687.5 & 2975.0 & 3350.0 & 3550.0 & 4025.0 & 4500.0 & \textit{5150.0} & {} & {} \\
 \hline
\end{tabular}
\end{center}
\caption{SDSS BOSS wavelength bands considered in this study. Italicized bands are rejected by our algorithm presented in Section \ref{sec:outliers}, and the band in boldface is chosen to be the proxy for the driving light curve as described in Section \ref{sec:driving_lc}. \tanote{All other wavelength bands are fitted by our method in Section \ref{sec:model_fitting}.}}
\label{table:wavelength_bands}
\end{table*}

We first subtracted the host-galaxy spectrum using the template found from spectral decomposition by \citet{shen2015msigma}. The host-galaxy contribution occasionally exceeds that of the AGN for rest frame $\lambda \gtrsim 4000$\AA\, and luminosities $\nu L_\nu \lesssim 10^{44}$ erg s$^{-1}$. Finally, we converted the observed-frame flux density to the emitted monochromatic luminosity $\nu L_\nu$ assuming luminosity distances based on a WMAP 9 year cosmology \citep{hinshaw2013} as implemented in \texttt{astropy} \citep{astropy} ($H_0 = 69.3\, \rm km\, \rm s^{-1} \, \rm Mpc^{-1}$, $\Omega_{\rm M} = 0.287$, $\Omega_\Lambda = 0.713$).

\section{Methods} \label{sec:methods}

The general procedure adopted for model fitting is as follows:

\begin{enumerate}
    \item We infer a common fractional uncertainty in luminosity for each object by analyzing the high-cadence 2014 data as described in Section \ref{sec:unc_anal}.
    \item We reject outliers as described in Section \ref{sec:outliers} based on a method that sets a minimum timescale for intrinsic quasar variability and rejects $> 3 \sigma$ variability observed on shorter timescales. We neglect wavelength bands with poor data quality ($> 25\%$ of data points rejected by our method) and those near the edges of the BOSS Spectrograph wavelength band. The wavelength bands considered for each object are presented in Table \ref{table:wavelength_bands} with rejected bands shown in italics.
    \item To obtain an approximate driving light curve, we interpolate the bluest light curve which is then scaled with parameters accounting for the light travel time-delay ($t_0$), difference in mean luminosity ($\overline{L_C}$), and geometric dilution of variability ($s$) as described in Section \ref{sec:driving_lc}. The wavelength band chosen as the proxy for the driving light curve is shown in boldface in Table \ref{table:wavelength_bands}.
    \item We consider reprocessing in a hemisphere geometry and a thick-disk geometry and obtain analytic light curves in particular wavelength bands by computing the appropriate geometric time-delay, effective temperature of the reprocessing surface, and corresponding spectra assuming a blackbody distribution (described in Section \ref{sec:quant}).
    \item We perform a simultaneous minimum-$\chi^2$ fit of all optical bands, excluding the interpolated wavelength band, in each reprocessing model. We analyze the best fit light curves, the fit parameter values, and corresponding reduced minimum-$\chi^2$ value (see Section \ref{sec:model_fitting}).
\end{enumerate}

We adopt the fiducial black-hole mass based on H$\beta$ \citep{grier2017}, C IV \citep{grier2019}, or Mg II \citep{shen2011, homayouni2020} line-width measurement where available in this order of preference, and we use the improved systemic redshift given by \citet{shen2019}. \tanote{Both of these are listed in Table \ref{table:rmid_sample} along with the point-spread function (PSF) magnitude in the SDSS u-band.} We use rest-frame monochromatic luminosities with host-galaxy light subtracted as described above, but consider the host-galaxy contribution when determining the appropriate uncertainty on data points. In particular, we calculate the uncertainty on each data point as a common fractional uncertainty on the total observed light (from both the quasar and the host-galaxy). We considered additional factors of reddening from the host-galaxy and the intergalactic medium, but our results showed that very large reddening of $E(B-V) \sim 0.1$--$0.3$ mag is required to make significant changes to the model fit results. The light curves are converted to that of the quasars' rest frame.

\newpage

\subsection{Uncertainty Analysis} \label{sec:unc_anal}

For each object, we estimate the luminosity uncertainty in each band empirically by using the high-cadence data from 2014. While the signal-to-noise of the observations is significantly higher than our empirical estimate, we take this approach in order to account for the systematic night-to-night differences due to calibration issues or atmospheric conditions. We fit a quadratic function to the high-cadence data to capture any intrinsic quasar variability during this time period, and assume that the remaining scatter is due to random noise independent of the intrinsic variability. Although the number of data points from 2014 is relatively small ($N = 32$), the residuals are distributed normally upon common-sense inspection and we deem it sufficient to fit a Gaussian to infer a characteristic uncertainty, $\sigma_\lambda$, from this fit. From the previously obtained quadratic fit, we also compute an average luminosity, $\overline{L_\lambda}$, over this time period. The quantity $\sigma_\lambda / \overline{L_\lambda}$ is the fractional luminosity uncertainty for that particular wavelength band.

We take the median of these fractional uncertainties to be the characteristic fractional uncertainty, $\delta_f$, for that quasar. For this analysis, we do not take into account the wavelength bands with recorded host-galaxy data, since we want to isolate the uncertainty on quasar light measurements. The resultant fractional uncertainties were in the range 6--29\%; in the rare cases where this empirically-measured fractional uncertainty was below $10$\%, we set $\delta_f = 10$\% as a baseline fractional uncertainty for our analysis. We note that this is a much more conservative uncertainty estimate than the $\sim 5 \%$ spectrophotometric accuracy reported in \citet{shen2015}, but since our model comparison only depends on relative $\chi^2$, our general conclusions are unaffected by our choice of fractional uncertainty. We apply this characteristic uncertainty as a common fractional uncertainty on all data points, so the uncertainty on each data point is calculated as $\delta_f \times (\mathrm{quasar \ light} + \mathrm{host \ galaxy \ light})$ across all wavelengths.

\subsection{Outlier Rejection and Wavelength Band Selection} \label{sec:outliers}

In this section, we describe an outlier-rejection method that accounts for the timescales over which intrinsic variability is expected to occur. In order to remove extreme deviations caused by night-to-night observational differences, we neglect extreme variability shorter than a month ($= 30 \ \mathrm{days}$), with time measured in the rest frame of the quasar. For each data point in a particular wavelength band, we take the median of all data points within a month and reject the data point if it lies outside $3 \sigma$ of the median. Our assumption here is that $> 3 \sigma$ quasar variability within 30 days is unlikely, and we do this in part to prevent the model from over-fitting the high cadence data of 2014. Generally, a few percent of data points, and up to $\sim 10 \%$ for bands closer to the BOSS Spectrograph edges, are rejected by this method.

An entire wavelength band of data is rejected when more than a quarter of the data points are removed via the above outlier rejection method. In addition, we exclude any wavelength bands that fall outside of the range $4100$-$9000 \ \rm{\AA}$ in the observed reference frame since spectrophotometry is considerably worse near the blue ($3600$ \AA) and red ($10400$ \AA) edges of the spectrograph \citep{sun2015}. The wavelength bands considered are shown in Table \ref{table:wavelength_bands}, and the ones that are rejected are italicized.

\subsection{Estimating the Driving Light Curve}
\label{sec:driving_lc}

After performing outlier and wavelength-band rejection, we interpolate the bluest available light curve to obtain a base template for the driving light curve. \tanote{For simplicity, we use linear interpolation. Data in all wavelength bands are collected at the same cadence, so we deem it sufficient to study their correlated variability.} The observed bluest continuum is only a proxy for the driving continuum, which is unobservable due to the Lyman limit of the host-galaxy. XMM-Newton \citep{xmm} data are available for some of our objects \citep{Liu2020}, but the data are not collected with sufficient cadence to examine whether the X-ray light curve variability is correlated with the optical variability. Therefore, we opt to interpolate the bluest optical band available. The bluest available light curve chosen as a proxy for the driving light curve is shown in boldface for each object in Table \ref{table:wavelength_bands}.

Our reprocessing model parameterizes the potential differences between the driving light curve and the bluest observed light curve. First, a geometric time-delay is expected between the driving light curve and the interpolated blue light curve. We add a parameter, $t_0$, which is simply a horizontal shift to the interpolated light curve, to account for this effect. We also expect the mean luminosity to differ between the bluest optical band and the driving light curve, so we re-normalize the interpolated light curve to the mean driving luminosity, $\overline{L_C}$. Finally, quasars are more variable in the UV compared to the optical \citep[e.g.,][]{Wilhite05, macleod2016}. A vertical stretch parameter, $s$, is introduced which scales the light curve's deviation from its mean luminosity.

\subsection{Quantitative Models}

\label{sec:quant}

We now explore the two reprocessing geometries for our model fitting, namely the hemisphere and the thick-disk model. In each case, we include the appropriate geometric parameters to analytically predict the obtained light curve at a particular wavelength band based on our estimated driving light curve.

\subsubsection{Hemisphere Model}

\begin{figure}[t!]
\centering
\includegraphics[width=\linewidth]{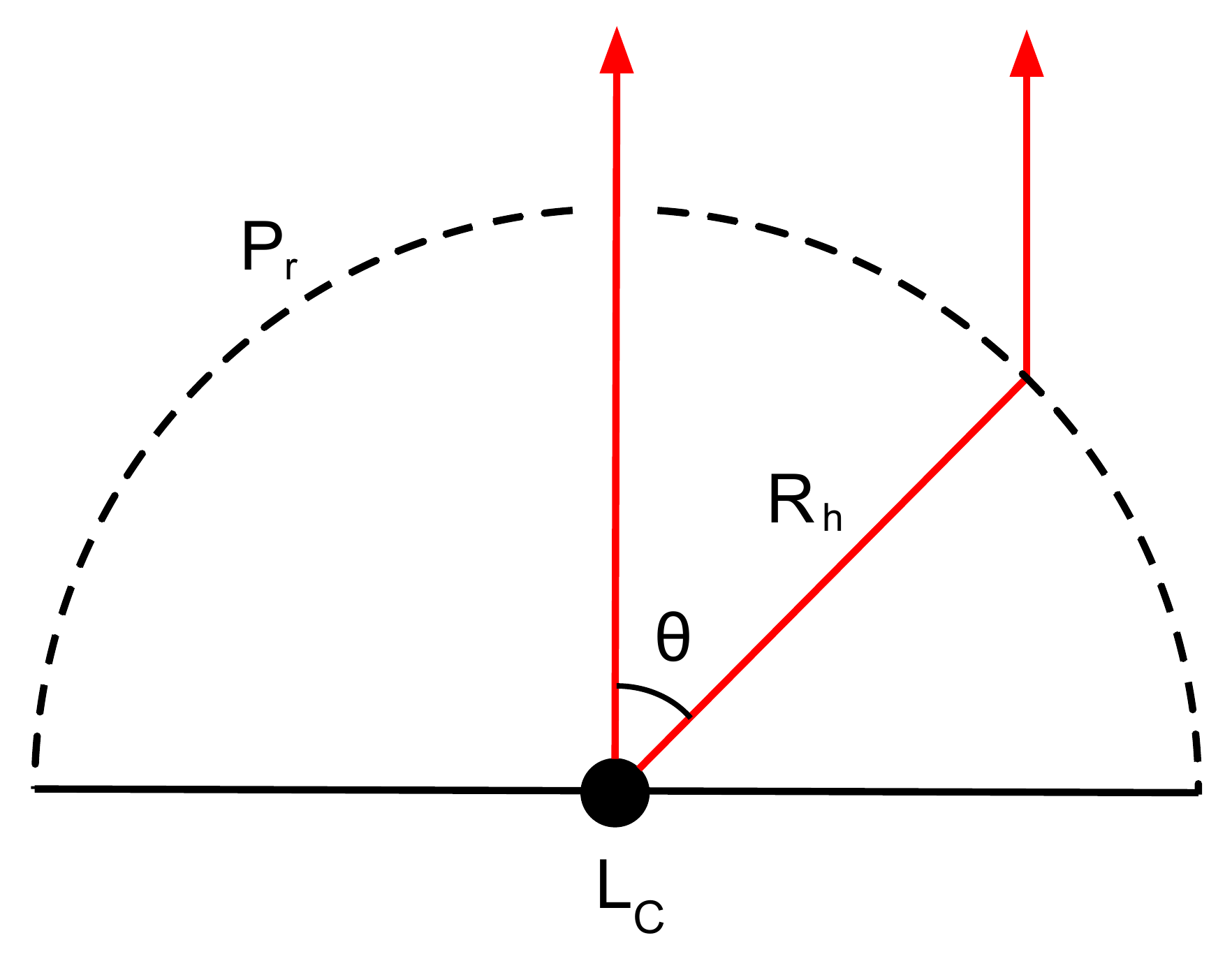}
\caption{A schematic diagram of the hemisphere geometry with two light rays. The central source is located at $L_C$ with a single reprocessing hemisphere at radius $R_h$. There is a probability, $P_r$, of light being reprocessed at this radius. The light ray on the right-hand side is initially emitted at an angle $\theta$ away from the observer and is reprocessed by the hemisphere. \tanote{The associated light travel time-delay is described by Equation \ref{eqn:time_delay_hemisphere}.} The line of sight to the observer (upward in the diagram) is assumed to be unobscured. \tanote{Face-on geometry is assumed.}}
\label{fig:hemisphere}
\end{figure}

We first consider the hemisphere model, where the geometry is a single reprocessing hemisphere at a certain radius from the central source. A schematic diagram of this reprocessing geometry is shown in Figure \ref{fig:hemisphere}. The driving light curve is emitted from the center labeled $L_C$ and is reprocessed by a hemisphere at radius $R_h$ with some probability $P_r$. The observer is assumed to be upward in the diagram. In this simplified model, $P_r$ directly corresponds to the covering factor of this geometry. If we set $P_r \sim 1$, all of the driving light curve is captured by the hemisphere before reaching the observer and the entire optical spectrum is powered by reprocessing. In reality, this geometry is an approximation for any out-of-plane, optically thick material caused by outflows/winds, warped/tilted disks, etc. where we have assumed that our view is unobstructed. The line-of-sight allowed obstruction due to the reprocessing structure is significantly constrained by X-ray absorption and other observations of the accretion disk. In this study, we assume that the emission from the accretion disk is unobscured even when $P_r \sim 1$. \tanote{We assume a face-on geometry for simplicity, but since $L_C$ is a point-source at the center of the hemisphere, the reprocessing structure has no inclination-dependence. Inclination will only alter the contribution from the accretion disk itself.}

\tanote{We follow the method of \citet{shappee2014}, and} first calculate the expected geometric time-delay as a function of the radius of the hemisphere, $R_h$, and the polar angle, $\theta$, of the emitted light ray. From the schematic in Figure \ref{fig:hemisphere}, one can determine using geometry that the light travel time-delay is given by

\begin{equation}
    \tau = \frac{R_h(1 - \mu)}{c} \ ,
\label{eqn:time_delay_hemisphere}
\end{equation}
    
\noindent where $\mu \equiv \cos \theta$. If we consider a variable luminosity source $L_C (t)$ at the center, we can calculate the effective temperature of the hemisphere as a function of time and the above geometric parameters. Since the luminosity is reprocessed at radius $R_h$ and delayed by a time $\tau$,
    
\begin{equation}
    \sigma T^4 = P_r \cdot \frac{L_C (t - \tau)}{4 \pi R_h^2} \ ,
\label{eqn:hemisphere_T}
\end{equation}

\noindent where $P_r$ is the probability of the light being reprocessed at radius $R_h$, and $\tau$ is given by Equation \ref{eqn:time_delay_hemisphere}. Finally, we assume a blackbody distribution with the above effective temperature for the reprocessed light, producing the spectrum
    
\begin{equation}
    \nu L_\nu = \frac{16 \pi^2 h \nu^4}{c^2} \int_0^1 \frac{R_h^2 \mu \ d \mu}{\exp(h \nu/k T)-1} \ ,
\label{eqn:hemisphere_nuLnu}
\end{equation}

\noindent where $\nu$ is the frequency of light and $T$ is given by Equation \ref{eqn:hemisphere_T}.

We also include the contribution from accretion parameterized by the accretion luminosity, $L_A$. As a first-approximation, we apply the standard thin disk formula where the effective temperature is given by
    
\begin{equation}
    \sigma T^4 = \frac{3 L_A R_{\mathrm{in}}}{2 \pi R^3} \left[ 1 - \left( \frac{R_{\mathrm{in}}}{R} \right)^{1/2} \right] \ ,
\label{eqn:thin_disk_T}
\end{equation}

\noindent where $R_{\mathrm{in}} \equiv 6 \ R_g$ (gravitational radius). The contribution to the spectrum from accretion is then
    
\begin{equation}
    \nu L_{\nu} = \frac{8 \pi^2 h \nu^4}{c^2} \int_{R_{\mathrm{in}}}^{\infty} \frac{R \ dR}{\exp(h \nu/k T) - 1} \ .
\label{eqn:thin_disk_nu_Lnu}
\end{equation}

\noindent Equation \ref{eqn:hemisphere_nuLnu} combined with Equation \ref{eqn:thin_disk_nu_Lnu} generates the expected total spectra for the observed optical bands based on our estimated driving light curve and parameters $R_h$, $P_r$, and $L_A$. Here, we set an arbitrary minimum temperature of $T = 300 \ \mathrm{K}$ to avoid numerical issues. Contributions to the UV/optical spectrum at temperatures this cold are negligible. The minimum-$\chi^2$ fit can then be calculated by taking into account $R_h$, $P_r$, and $L_A$ as well as the three parameters required for producing the driving light curve: $t_0$, $\overline{L_C}$, and $s$.

\subsubsection{Thick-Disk Model}

\begin{figure}[t!]
\centering
\includegraphics[width=\linewidth]{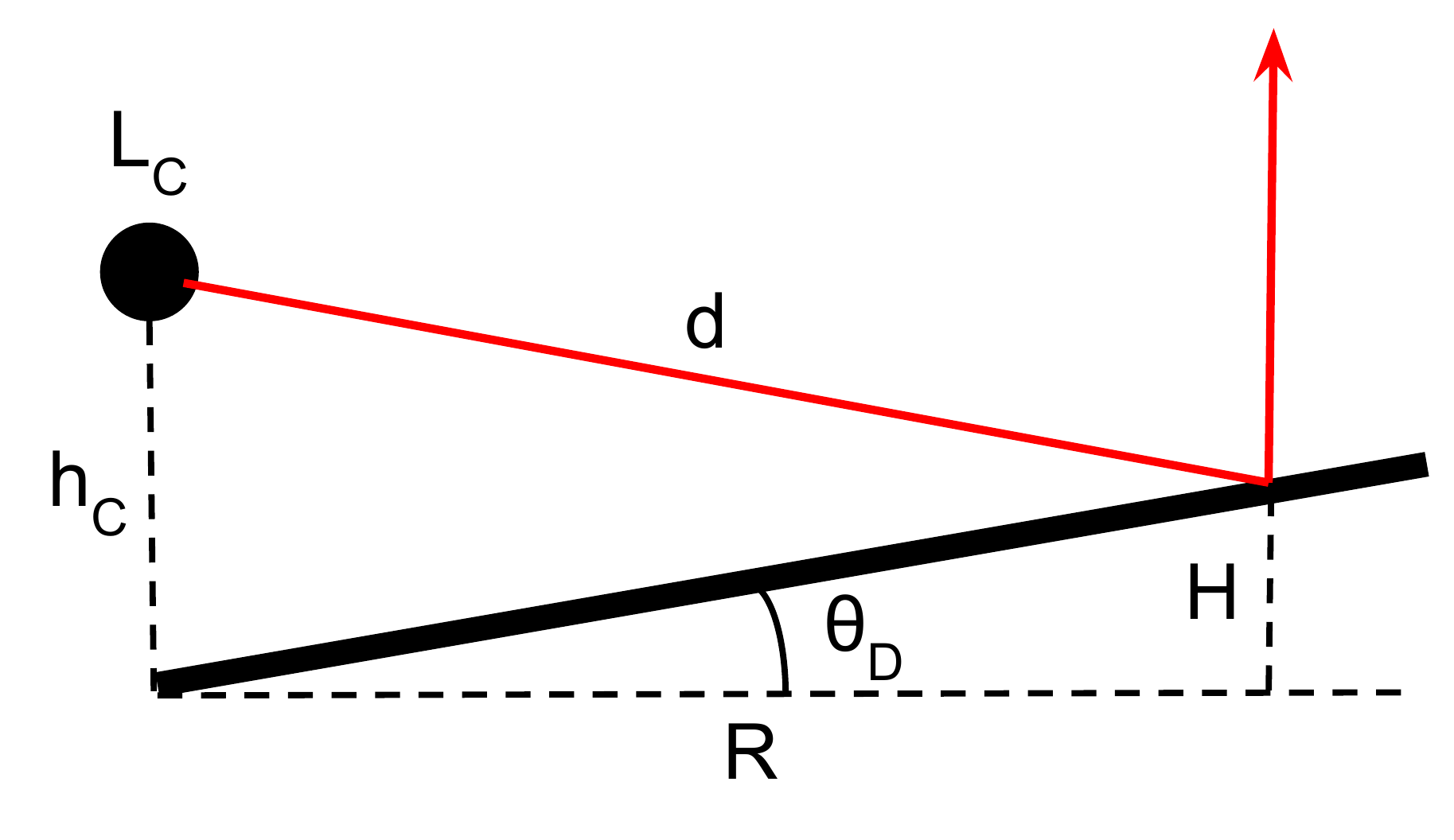}
\caption{A schematic diagram of the thick-disk model with an example reprocessed light ray. The central source $L_C$ is at some height $h_C$ above the disk, and the disk is elevated by an angle $\theta_D$. The point at which the light is reprocessed is a radius $R$ out from the center of the disk, correspondingly elevated by a height $H$, and a distance $d$ from the central source. \tanote{The associated light travel time-delay is given by Equation \ref{eqn:h_r_disk_time_delay}. Face-on geometry is assumed.}}
\label{fig:H_R_disk}
\end{figure}

Next, we consider the thick-disk model, an accretion disk where the height-to-radius ratio, $H/R$, is non-negligible. A schematic of a cross-section of this thick disk is shown in Figure \ref{fig:H_R_disk}. We once again assume that the driving light curve is emitted from the point on the diagram labeled $L_C$, but in this model the central source is assumed to be at some height, $h_C$. This simple ``lamp post'' model is inspired by \citet{cackett2007} and \citet{shappee2014} with the only difference being that the disk is elevated. X-ray emission from quasars has a half-light radius $\sim 10 \ R_g$ \citep{mosquera2013}, which is small enough to treat the X-ray emission as a point source \citep{shappee2014}. We assume that $H/R$ is constant throughout the disk. In this case, not all of the light emitted from $L_C$ can be reprocessed before emerging from the quasar. The covering factor monotonically increases as we raise the disk, so the model fit is expected to choose an appropriate $H/R$ that suits the geometric dilution observed in our light curves. It is also important to note that we leave $H/R$ as a free parameter, so our model also includes reprocessing in a thin disk in the limit that $H/R = 0$. \tanote{We once again assume a face-on geometry here, since reprocessing in a significantly elevated disk is possible only for low inclinations.}

As before, we first calculate the geometric time-delay. We can once again use geometry to find that the light travel time-delay is

\begin{equation}
    \tau = \frac{d + (h_C - R \sin \theta_D)}{c} \ ,
\label{eqn:h_r_disk_time_delay}
\end{equation}
    
\noindent where $d = \sqrt{(h_C - R \sin \theta_D)^2 + (R \cos \theta_D)^2}$. We parameterize the contribution from accretion by the accretion luminosity, $L_A$. In addition, we include the reprocessing contribution term obtained from our geometry,
    
\begin{equation}
\begin{split}
    \sigma T^4 = & \frac{3 L_A R_{\mathrm{in}}}{2 \pi R^3} \left[ 1 - \left( \frac{R_{\mathrm{in}}}{R} \right)^{1/2} \right] \\ & + \frac{L_C (t - \tau)}{4 \pi d^2} \frac{h_C \cos \theta_D}{d} 
\end{split}
\label{eqn:h_r_disk_T}
\end{equation}

\noindent where $R_{\mathrm{in}} \equiv 6 \ R_g$, $\tau$ is given by Equation \ref{eqn:h_r_disk_time_delay}, and $d$ is as before. Finally, we obtain the total spectra assuming a blackbody distribution and integrating over $R$,
    
\begin{equation}
    \nu L_{\nu} = \frac{8 \pi^2 h \nu^4 \cos \theta_D}{c^2} \int_{R_{\mathrm{in}}}^{\infty} \frac{R \ dR}{\exp(h \nu/k T) - 1} \ .
\label{eqn:h_r_disk_nu_Lnu}
\end{equation}

\noindent Equation \ref{eqn:h_r_disk_nu_Lnu} determines the expected spectral appearance at a particular wavelength band by assuming geometric parameters $h_C$, $\theta_D$, the accretion luminosity contribution $L_A$, and the three parameters required for producing the driving light curve: $t_0$, $\overline{L_C}$, and $s$. Once again, we set a minimum temperature of $T = 300 \ \mathrm{K}$.

\tanote{We note that this method of quantifying the time-delay in each reprocessing geometry and obtaining the spectrum in each wavelength band is equivalent to the notion of a transfer function \citep{peterson1993, collier1998} which characterizes the response of the reprocessor to a $\delta$-function light source. The transfer function is given by}

\begin{equation}
\Psi_\nu(t, \lambda) = \int_{R_{\rm{in}}}^{R_{\rm{out}}} \frac{\partial B_\nu}{\partial T} \frac{\partial T}{\partial L_C} \times \delta(t - \tau) \ d \Omega \ ,
\end{equation}

\begin{figure}[t!]
\centering
\includegraphics[width=\linewidth]{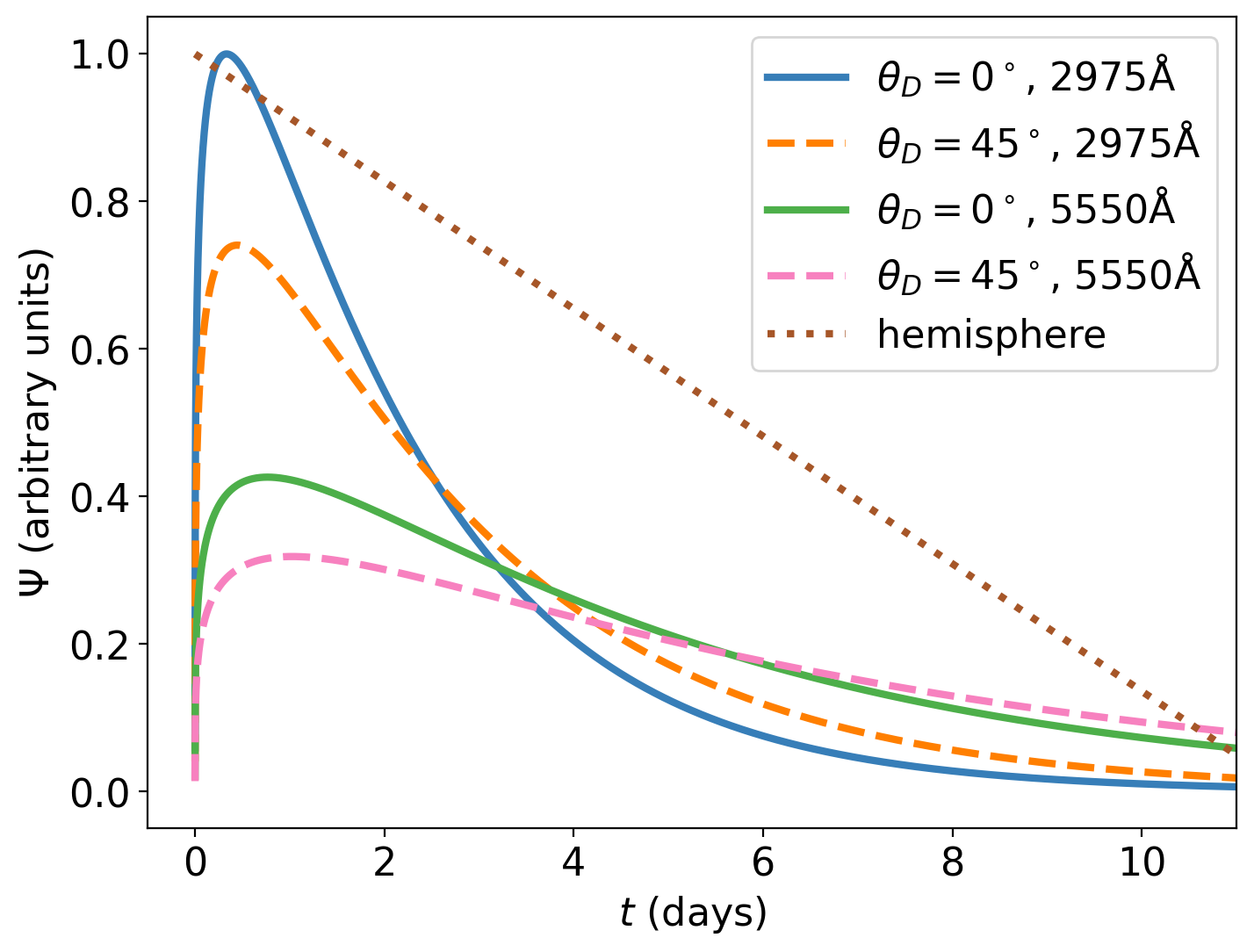}
\caption{\tanote{Sample transfer functions at two wavelengths ($2975 \rm{\AA}$ and $5550 \rm{\AA}$) for the disk model (with $\theta_D = 0^\circ$ and $\theta_D = 45^\circ$), and the hemisphere model. The disk model when $\theta_D = 0^\circ$ is consistent with the standard thin disk transfer function \citep[e.g.,][]{cackett2007}. The transfer function retains the same shape but is wider at the same wavelength for an elevated disk ($\theta_D = 45^\circ$). The hemisphere transfer function is a linear decay with no wavelength-dependence.}}
\label{fig:trans_func}
\end{figure}

\noindent \tanote{where time-delays in our two geometries are given by Equations \ref{eqn:time_delay_hemisphere} and \ref{eqn:h_r_disk_time_delay}. As a proof of concept, we plot sample transfer functions obtained using our quantitative models in Figure \ref{fig:trans_func}. We show transfer functions for two wavelengths, $2975 \rm{\AA}$ and $5550 \rm{\AA}$, with the thin ($\theta_D = 0^\circ$) and thick ($\theta_D = 45^\circ$) disk, and we additionally show the transfer function for the hemisphere model. The disk model when $\theta_D = 0$ recovers the standard thin disk transfer function \citep[e.g.,][]{cackett2007}. The thick disk has the same shape, but the time-delay has a dependence on $\theta_D$ given by $\tau \sim \lambda^{4/3} (1 - \sin\theta_D)$. For the same wavelength, an elevated disk will have a wider transfer function than a thin disk. For the hemisphere model, the transfer function is a linear decay with no wavelength-dependence. We note that a more realistic hemisphere-like reprocessor could be multi-temperature and/or radially extended in which case the time-delay would be wavelength-dependent. Our analysis, however, only considers a single reprocessing hemisphere.}

\subsection{Model Fitting}
\label{sec:model_fitting}

\begin{figure*}[t!]
\begin{tabular}{cc}
\includegraphics[width=\columnwidth]{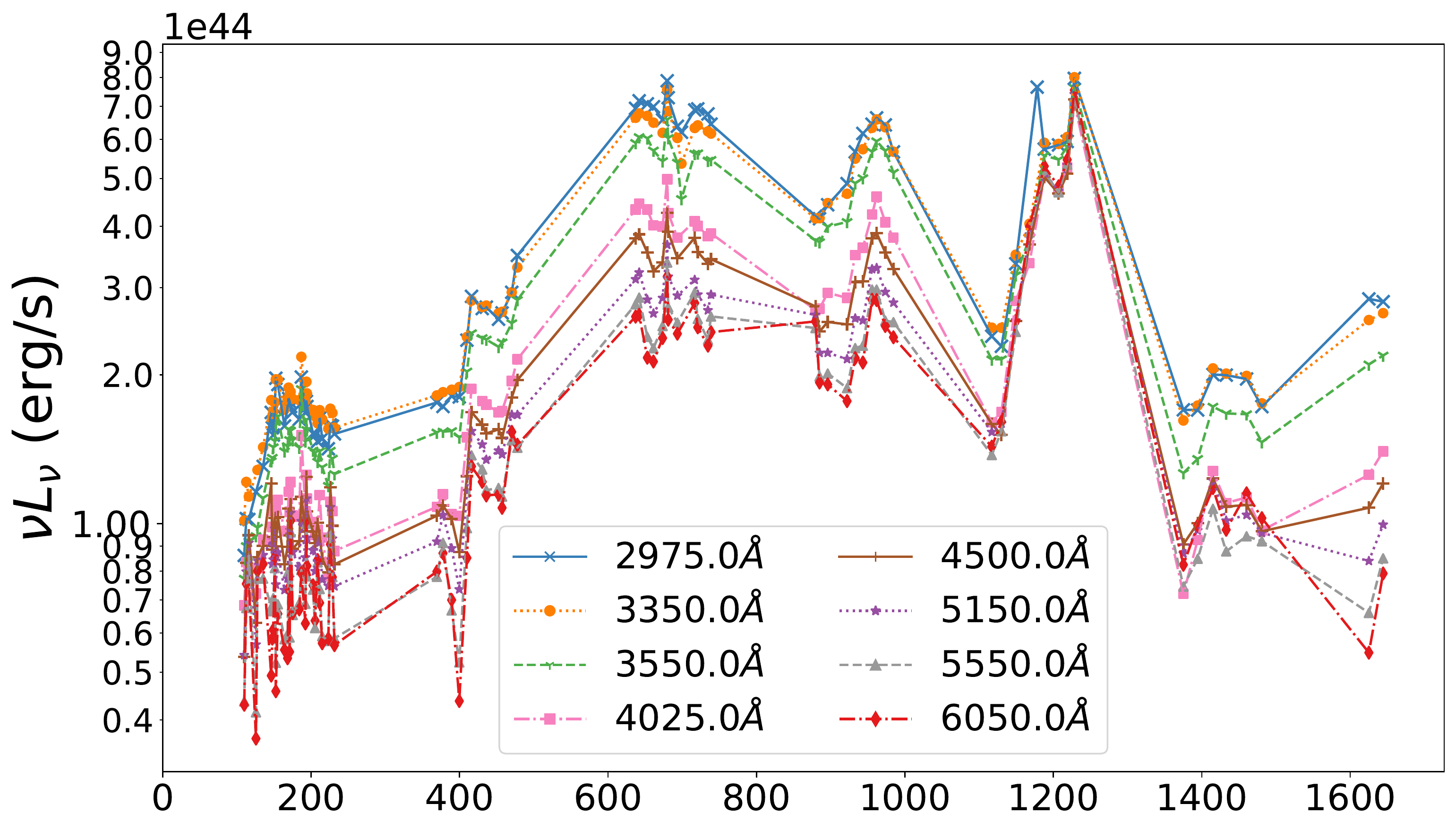} & \includegraphics[width=0.97\columnwidth]{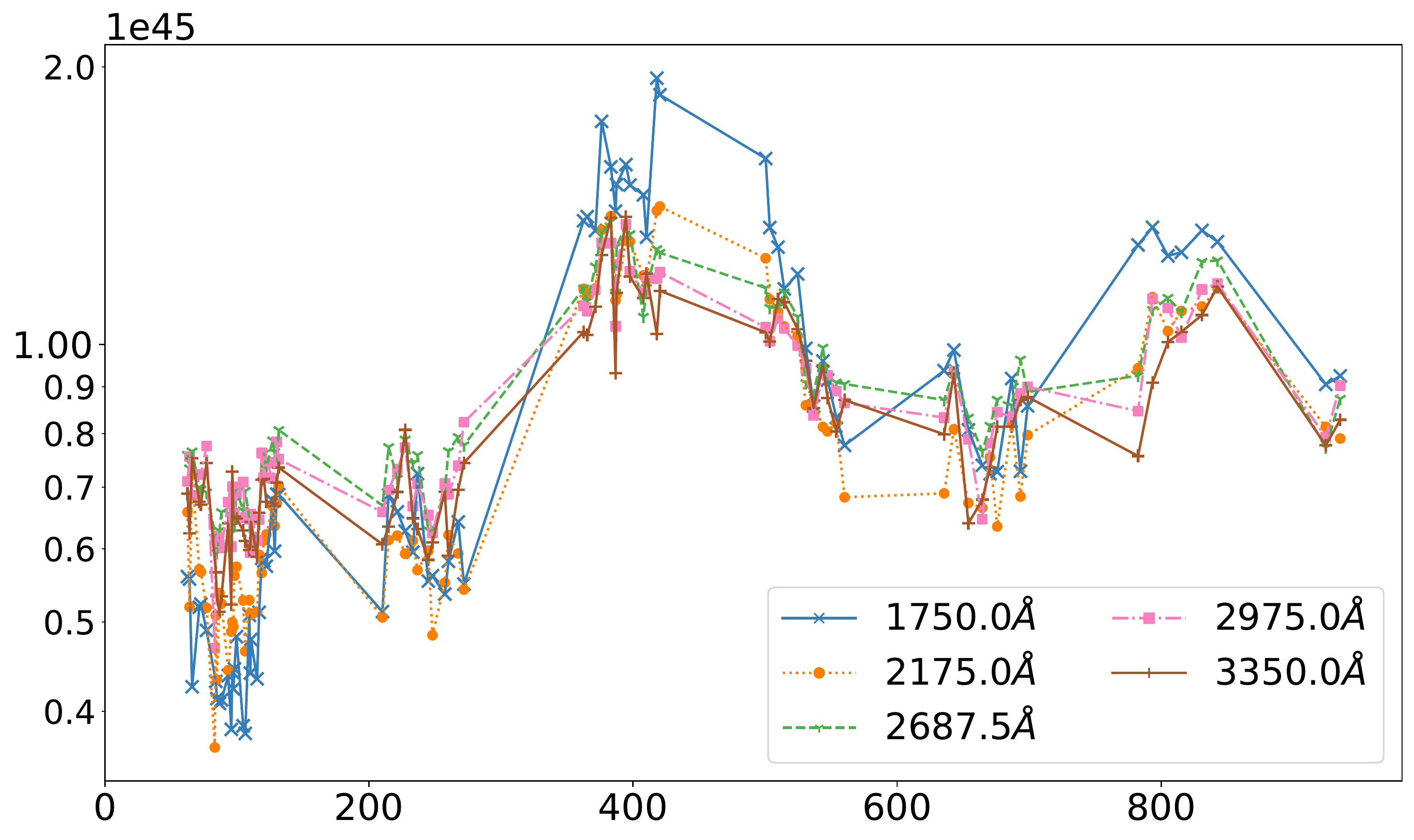} \\
(a) RM 17 & (b) RM 194 \\
\includegraphics[width=\columnwidth]{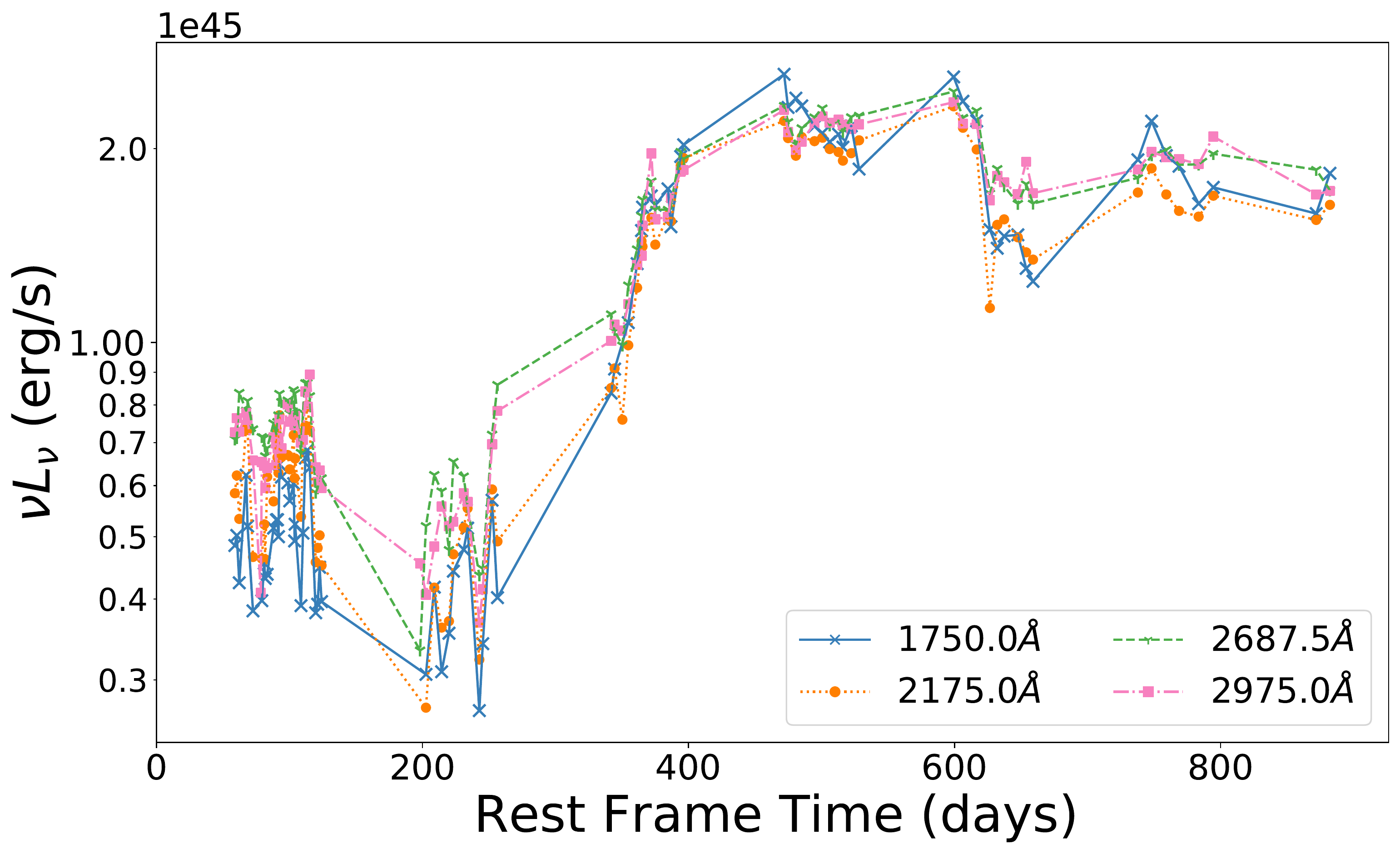} & \includegraphics[width=0.95\columnwidth]{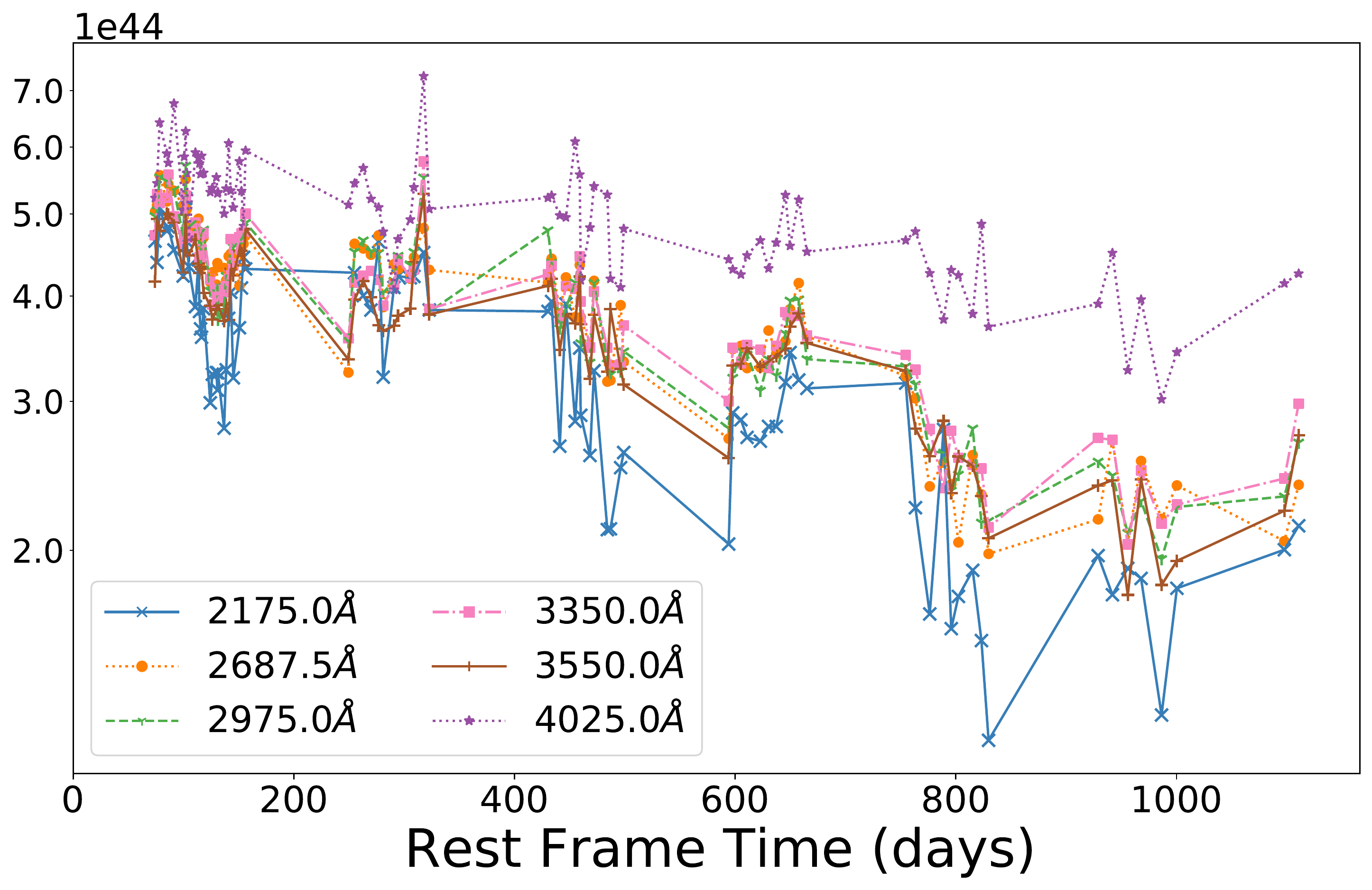} \\
(c) RM 32 & (d) RM 105 \\
\end{tabular}
\caption{Optical light curves constructed from SDSS spectra of objects in our case studies: (a) RM 17, (b) RM 194, (c) RM 32, and (d) RM 105. These four examples highlight the broad range of color/variability behavior we see in the full sample. In each panel, we show the spectrum, $\nu L_{\nu}$, for each available optical band over rest-frame time in days.}
\label{fig:case_study_lc}
\end{figure*}

We fit Equation \ref{eqn:hemisphere_nuLnu} combined with Equation \ref{eqn:thin_disk_nu_Lnu} (for the hemisphere), and Equation \ref{eqn:h_r_disk_nu_Lnu} (for the thick disk) to the observed light curves, simultaneously fitting all available optical bands (excluding the bluest band that was chosen as a proxy for the driving light curve) and finding the best-fit parameters that minimize $\chi^2$. We use the \texttt{scipy.optimize.curve\_fit} function which uses the Trust Region Reflective algorithm for bound problems. While we evaluate goodness-of-fit based on $\chi^2$ without any scaling, the uncertainties on parameter values reported in Tables \ref{table:hemisphere_fit_param_vals} and \ref{table:disk_fit_param_vals} are based on scaling reduced $\chi^2$ to 1. We deem a simple minimum-$\chi^2$ fit to be sufficient for model comparison between the hemisphere and thick-disk geometries, since each requires the same number of parameters (6). In this case, comparisons using the Akaike Information Criterion (AIC) or the Bayesian Information Criterion would be equivalent. As a test case, we performed reprocessing model fits on NGC 2617 with these two geometries following \citet{shappee2014}. The model preferred by our minimum-$\chi^2$ analysis was either a thin disk, consistent with \citet{shappee2014}, or a hemisphere with a very low covering factor ($P_r \sim 10^{-2}$) which is effectively a similar geometry.

\section{Results} \label{sec:results}

\subsection{Classification Scheme}

While our sample of 17 hypervariable quasars exhibits a broad range of color/variability behavior, all but one object are successfully fit by either the hemisphere model or the thick-disk model. We explore a few of these successes (and the one failure) in case studies: we discuss RM 17 (Section \ref{sec:case_study_rm17}) to make connections to previous work \citep{dexter2019}, RM 194 and 32 (Sections \ref{sec:case_study_rm194} and \ref{sec:case_study_rm32}) to analyze how the color evolution of the optical light curves affects the selected covering factor, and RM 105 (Section \ref{sec:case_study_fail}) to describe the shortcomings of thermal reprocessing models. Figure \ref{fig:case_study_lc} shows the optical light curves of these case study objects. Tables \ref{table:hemisphere_fit_param_vals} and \ref{table:disk_fit_param_vals} provide a full description of our 17 targets with best-fit parameters and resulting reduced minimum-$\chi^2$ values for each geometry. Table \ref{table:summary_fit} compares the reduced minimum-$\chi^2$ values and presents our list of likely classifications with corresponding $\Delta$AIC values. Our results indicate that a thin disk reprocessing model is disfavored for our sample of 17 objects. We note that RM 17 is the only radio-loud object \citep{shen2019} implying that phenomena such as beamed jet emission are unlikely to have affected the variability of the quasars in general. We discuss model fit results for each of the four objects in detail below.

\subsection{Case Study: RM 17} \label{sec:case_study_rm17}

\begin{figure*}[t!]
\centering
\includegraphics[width=\linewidth]{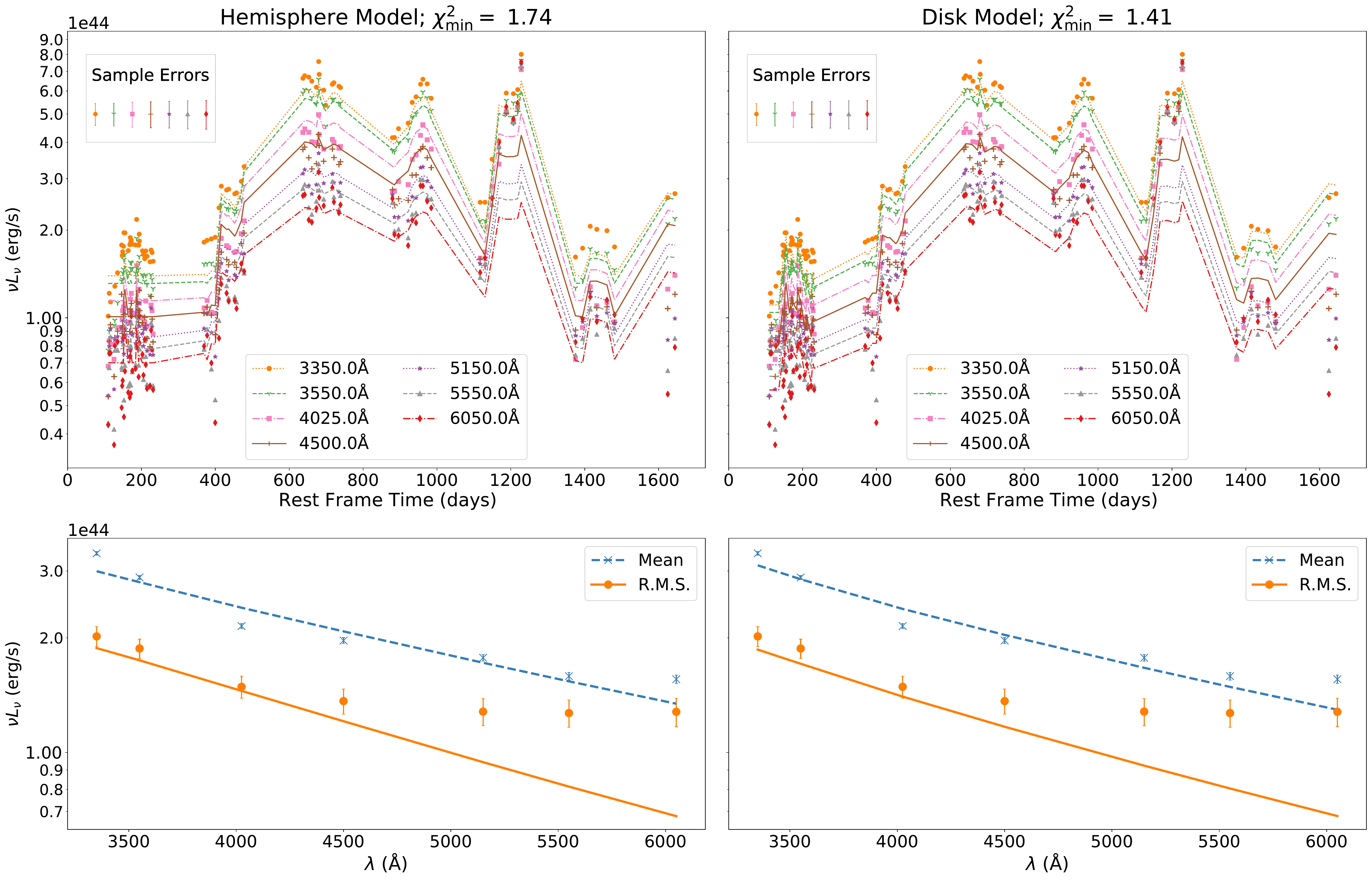}
\caption{(\textit{Top:}) Model fits (lines) compared to observed data (markers) for RM 17 in our two geometries: (\textit{left:}) hemisphere model and (\textit{right:}) disk model. \tanote{(\textit{Bottom:}) The mean (blue dashed lines with ``x'' markers) and rms (orange solid lines with ``o'' markers) spectra for RM 17 in each geometry. Once again the lines are model fits and markers are observed data.} The thick-disk geometry (right) is preferred for this object and reproduces most of the variability we see in each band except for the large spike in redder wavelength bands around $t \approx 1200$ days. Neither model is able to reproduce the abrupt change in the rms spectrum at $\lambda \sim 4500 \rm{\AA}$.}
\label{fig:rm17_fits}
\end{figure*}

RM 17, Figure \ref{fig:case_study_lc}(a), displays little to no color evolution in variability amplitude as we move from bluer to redder bands. There is little geometric dilution and the reddest bands still exhibit variability of factors $\sim 10$. \citet{dexter2019} demonstrated that this object is not well-fit by a thin-disk reprocessing model, likely because this object requires a much larger covering factor to describe its lack of color evolution in relative variability amplitude. Consequently, our expectation is that the analysis should show a preference for a thick disk or a hemisphere with a significant covering factor. The reduced minimum-$\chi^2$ fits for our two geometries are quoted in Tables \ref{table:hemisphere_fit_param_vals} and \ref{table:disk_fit_param_vals} and the corresponding model light curves compared to the observed data are shown in the top row of Figure \ref{fig:rm17_fits}. The disk model is preferred, but it selects a significant thickness of $H/R = 0.353$ and an accretion luminosity ($L_A$) substantially smaller than the reprocessed luminosity ($\overline{L_C}$). The results are consistent since a larger covering area increases the reprocessing efficiency. Furthermore, it selects a stretch factor of $s = 1.67$, suggesting that the driving light curve must be of much greater variability than what is exhibited by the bluest available optical light curve. As can be seen in the top right panel of Figure \ref{fig:rm17_fits}, the thick-disk model is able to mostly reproduce correct variability amplitudes in each of the wavelength bands. This model is a simplification and taking into account multiple reprocessing --- light that gets reprocessed more than once --- might do a better job (see Section \ref{sec:conclusion}).

\tanote{We also plot the mean and rms spectra of RM 17 comparing model fits to observed data in the bottom row of Figure \ref{fig:rm17_fits}. While the slope of the mean spectrum is reproduced fairly well by both models, the rms model spectra are unable to explain the abrupt change in the slope of the observed rms spectrum around $\lambda \sim 4500 \rm{\AA}$.} This is in large part due to the observations' large deviation from the model around $t \approx 1200$ days (discussed further in Section \ref{sec:case_study_fail}), especially at redder wavelength bands. To emphasize the magnitude of this deviation, we calculate the reduced minimum-$\chi^2$ values when we disregard data points with $1150 \ \rm{days} < t < 1250 \ \rm{days}$ which are much more reasonable. These values are also reported in Tables \ref{table:hemisphere_fit_param_vals} and \ref{table:disk_fit_param_vals} (in curly braces) in the final column. For this object specifically, one may suspect that the jet affected the hypervariable behavior since it is radio-loud, but the evidence for non-thermal jet emission is weak \citep{dexter2019}. The hemisphere model fit, Figure \ref{fig:rm17_fits} (left), selects the maximum covering factor ($P_r \sim 1$) with a significant contribution from accretion, somewhat contradictory to the thick disk result. Again, taking into account multiple reprocessing in our thick-disk model may alleviate some of this discrepancy. Nonetheless, geometries with a much larger covering factor are preferred in both cases. While a thin disk model was not successful for this object \citep{dexter2019}, the story is quite different when thick disks are considered.

\subsection{Case Study: RM 194} \label{sec:case_study_rm194}

\begin{figure*}
\centering
\includegraphics[width=\linewidth]{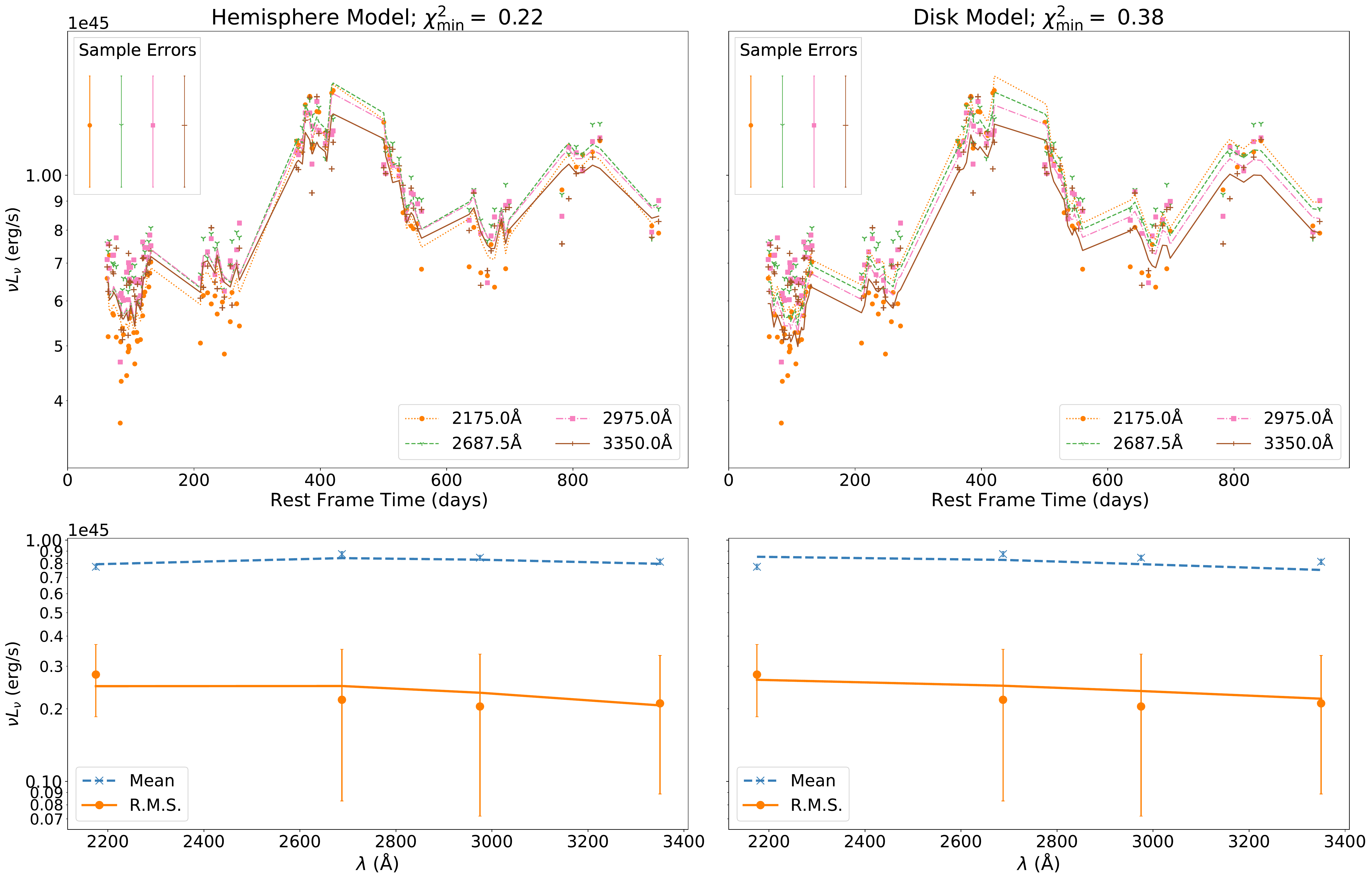}
\caption{(\textit{Top:}) Model fits (lines) compared to observed data (markers) for RM 194 in our two geometries: (\textit{left:}) hemisphere model and (\textit{right:}) disk model. \tanote{(\textit{Bottom:}) The mean (blue dashed lines with ``x'' markers) and rms (orange solid lines with ``o'' markers) spectra for RM 194 in each geometry. Once again the lines are model fits and markers are observed data.} While the hemisphere geometry (left) is preferred, both reprocessing models are successful for this object. Both models reproduce the mean and rms spectra well.}
\label{fig:rm194_fits}
\end{figure*}

\begin{figure*}
\centering
\includegraphics[width=\linewidth]{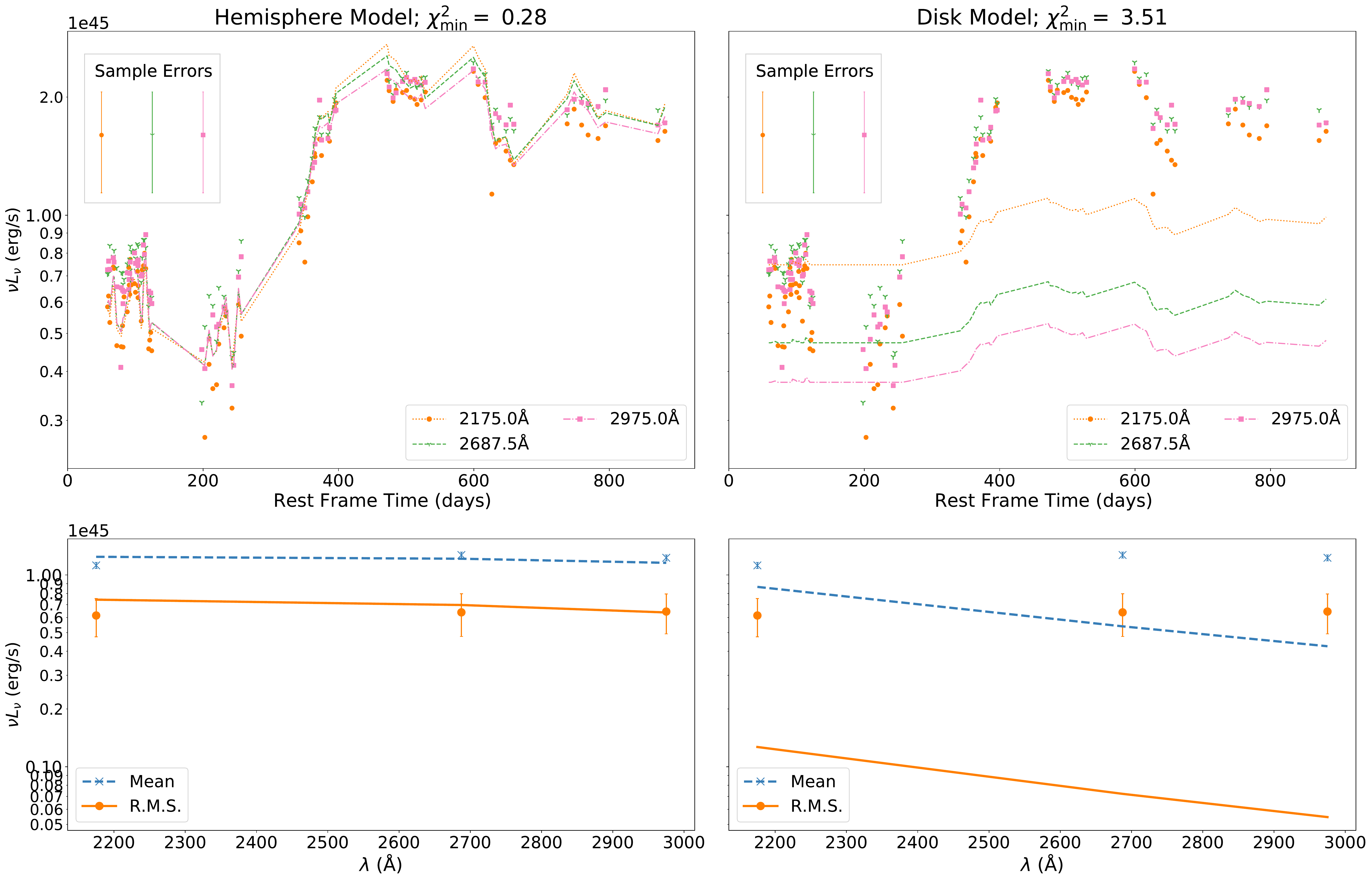}
\caption{(\textit{Top:}) Model fits (lines) compared to observed data (markers) for RM 32 in our two geometries: (\textit{left:}) hemisphere model and (\textit{right:}) disk model. \tanote{(\textit{Bottom:}) The mean (blue dashed lines with ``x'' markers) and rms (orange solid lines with ``o'' markers) spectra for RM 32 in each geometry. Once again the lines are model fits and markers are observed data.} The hemisphere geometry (left) is preferred and it is able to reproduce the light curve and mean and rms spectra fairly well. The thick-disk model is largely unsuccessful.}
\label{fig:rm32_fits}
\end{figure*}

Focusing our attention on RM 194, Figure \ref{fig:case_study_lc}(b), the bluest light curve displays significant variability but longer wavelength bands exhibit a much smaller variability amplitude. The short timescale variability in the redder bands is much closer to the $\sim 10 $--$ 20 \%$ seen in typical quasars. Our expectation is that a geometry with a smaller covering factor would be preferred here. The model fits for the hemisphere and disk geometries are detailed in Tables \ref{table:hemisphere_fit_param_vals} and \ref{table:disk_fit_param_vals} and the corresponding model light curves compared to the observed data are shown in the top row of Figure \ref{fig:rm194_fits}. The hemisphere model is preferred with a reduced covering factor ($P_r = 0.614$) with a significant contribution from the accretion luminosity, $L_A$. The disk geometry selects a thin disk ($H/R \sim 0$) which is also accompanied by a significant accretion luminosity term. Both models point to spectra which combine contributions from reprocessing as well as accretion, as opposed to spectra that are entirely powered by reprocessing. \tanote{The corresponding model and observed mean and rms spectra are shown in the bottom row of Figure \ref{fig:rm194_fits}. In both geometries, the model spectra largely capture the observed mean and rms spectra.}

\newpage

\subsection{Case Study: RM 32} \label{sec:case_study_rm32}

RM 32, Figure \ref{fig:case_study_lc}(c), exhibits little to no color evolution in variability amplitude. Our expectation suggests that this behavior requires a covering factor $\sim 1$, which should prefer a thick disk or a hemisphere. The minimum-$\chi^2$ fits for our two geometries are recorded in Tables \ref{table:hemisphere_fit_param_vals} and \ref{table:disk_fit_param_vals}, and the corresponding model light curves compared to the observed data are shown in the top row of Figure \ref{fig:rm32_fits}. The hemisphere model, Figure \ref{fig:rm32_fits} (left), is preferred with the maximum covering factor ($P_r \sim 1$). As can be seen in the figure, this model reproduces much of the overall optical variability successfully. The best-fit reprocessing hemisphere for RM 32 is located at $R = 720 \ R_s$ (Schwarzschild radius), which is farther from the black hole than where much of the optical emission is expected from an accretion disk. We base our disk size on the \citet{shakura1973} analytic result, shown to be in agreement (within 1.5 $\sigma$) with SDSS-RM optical disk size measurements \citep{homayouni2019}. This result is consistent with our simple physical picture that there is some reprocessing hemisphere that lies outside of our accretion structure. The disk fit, Figure \ref{fig:rm32_fits} (right), is largely unsuccessful. Due to geometric dilution, our model light curve is systematically ``undershooting'', indicating that this object is not well-modeled by a disk-like geometry. \tanote{The mean and rms spectra shown in the bottom row of Figure \ref{fig:rm32_fits} tell the same story: both the mean and rms spectra are well-fit by the hemisphere model but not the disk model.}

\subsection{Case Study: RM 105 and Other Failures of Reprocessing} \label{sec:case_study_fail}

\begin{figure*}
\centering
\includegraphics[width=\linewidth]{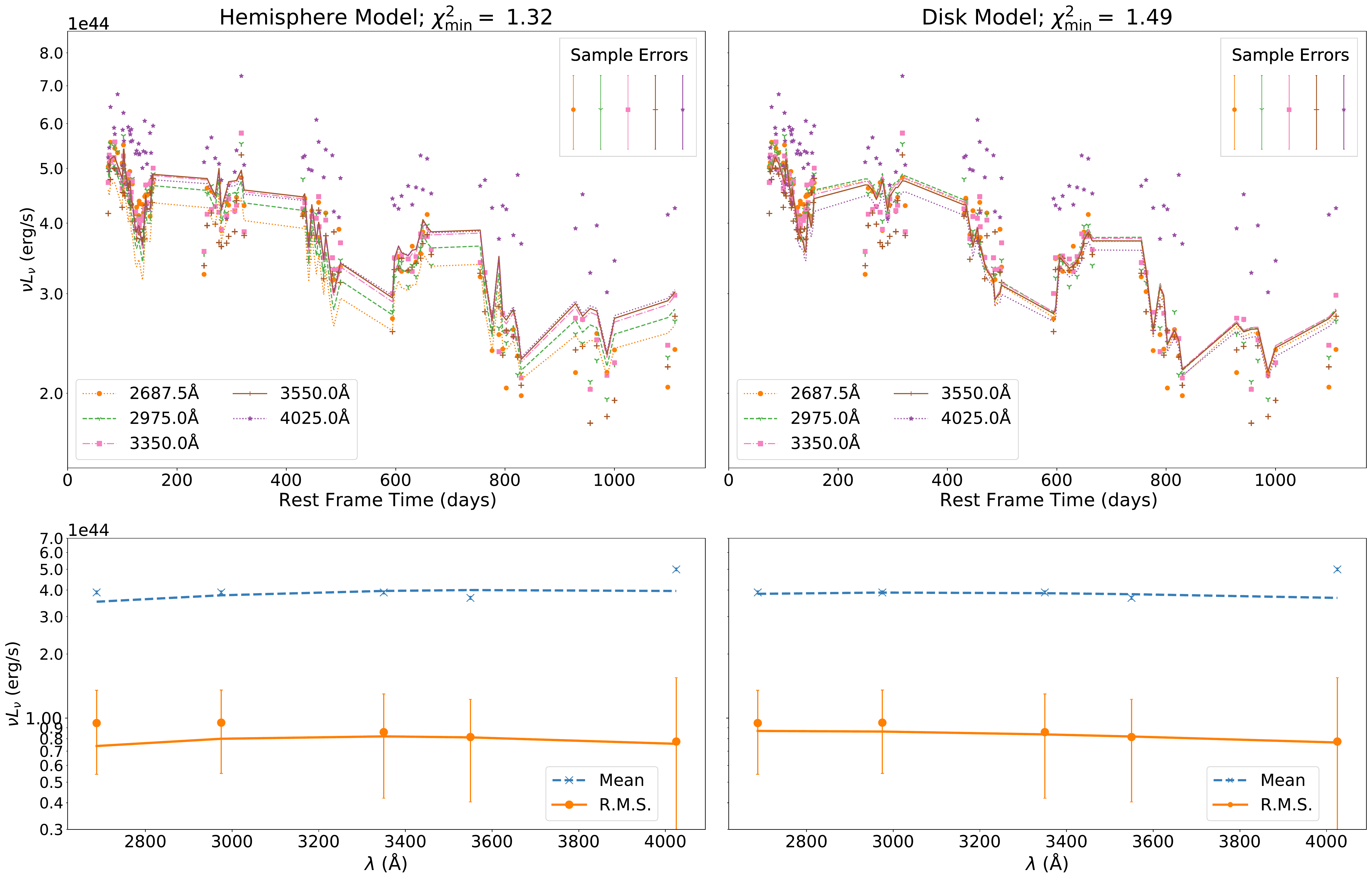}
\caption{(\textit{Top:}) Model fits (lines) compared to observed data (markers) for RM 105 in our two geometries: (\textit{left:}) hemisphere model and (\textit{right:}) disk model. \tanote{(\textit{Bottom:}) The mean (blue dashed lines with ``x'' markers) and rms (orange solid lines with ``o'' markers) spectra for RM 105 in each geometry. Once again the lines are model fits and markers are observed data.} Neither model is able to capture the color evolution, especially in the reddest wavelength band. The observed mean spectrum deviates significantly at the reddest wavelength band from both models.}
\label{fig:rm105_fits}
\end{figure*}

% \begin{figure*}
% \begin{tabular}{cc}
% \includegraphics[width=1.09\columnwidth]{RM105_hemisphere.pdf} & 
% \includegraphics[width=0.9\columnwidth]{RM105_HR.pdf} \\
% (a) Hemisphere geometry & (b) Disk geometry \\
% \end{tabular}
% \caption{Model fits (lines) compared to observed data (x) for RM 105 in our two geometries. Neither model captures the overall color evolution of variability we observe in this object.}
% \label{fig:rm105_fits}
% \end{figure*}

While most of the objects' light curves are well-modeled by reprocessing in either the hemisphere or thick-disk geometry, there remain objects where our quantitative models failed to characterize the variability amplitude in certain bands or times. For instance, RM 105, Figure \ref{fig:case_study_lc} (d), exhibits a higher variability amplitude and luminosity in the redder bands as compared to the bluer bands. This suggests that reprocessing alone cannot explain these optical spectra. Even with the maximum covering factor of 1, the variability amplitude in a redder wavelength band will not be able to exceed what is observed in the blue with just reprocessing. We quote the minimum-$\chi^2$ fit for RM 105 in each of our geometries in Tables \ref{table:hemisphere_fit_param_vals} and \ref{table:disk_fit_param_vals}, and the corresponding model light curves compared to the observed data are shown in the top row of Figure \ref{fig:rm105_fits}. Neither geometry is able to fit the optical light curves successfully. While the reddest band, $4025$ \AA, has the most striking discrepancy, the general amplitude pattern is not reproduced. The data reveal that the variability amplitude and luminosity increase in redder bands, but our reprocessing model does the opposite. No matter what the value of the covering factor is, our model always undershoots the observed luminosity and variability of the redder bands. While this result does not rule out the possibility that reprocessing contributes to the overall optical spectra for this object, there must exist some other physical mechanism, such as disk instability \citep[e.g.,][]{noda2018} or rapid inflow \citep[e.g.,][]{dexterbegelman2019}, by which variability amplitudes are increased in the redder bands. \tanote{We show the mean and rms spectra compared between model fits and observed data in the bottom row of Figure \ref{fig:rm105_fits}. The observed mean spectrum deviates significantly from both model spectra at the reddest wavelength.}

\begin{table*}[t!]
\begin{center}
\begin{tabular}{||c||c|c|c|c|c||c||} 
 \hline
 \multirow{2}{*}{RM-ID} & \multicolumn{6}{c||}{Hemisphere} \\
 {} & $P_r$ & $\overline{L_C} \ [10^{45} \ \rm{erg} \ \rm{s}^{-1}]$ & $L_A/\overline{L_C}$ & $R_h \ [R_s]$ & $s$ & $\chi_{\mathrm{min}}^2/N$ \\
 \midrule
 12 & 0.0619 (0.0282) & 10 (4.54) & 0.0331 (0.0170) & 526 (109) & 1 (0.0817) & 0.907 \\
 17 & $\sim$ 1 (14.4) & 0.613 (8.84) & 0.529 (7.63) & 52.7 (379) & 2 (0.0480) & 1.74 \{1.32\} \\
 32 & $\sim$ 1 (1.33) & 3.35 (4.46) & 0.736 (1.013) & 720 (472) & 1 (0.0216) & 0.284 \\
 \textit{105*} & \textit{0.609 (0.724)} & \textit{0.890 (1.06)} & \textit{0.207 (0.255)} & \textit{32.9 (19.4)} & \textit{1 (0.0966)} & \textit{1.32} \\
 112 & $\sim$ 1 (1.67) & 1.11 (1.85) & 0.303 (0.509) & 13.2 (11.0) & 1 (0.0720) & 1.03 \\
 143 & 0.230 (0.207) & 2.72 (2.46) & 0.207 (0.188) & 55.1 (24.7) & 2 (0.0824) & 0.805 \\
 160 & $\sim$ 1 (2.62) & 0.0947 (0.248) & 0.880 (2.30) & 107 (140) & 1 (0.506) & 1.34 \\
 194 & 0.614 (0.412) & 2.46 (1.66) & 0.182 (0.127) & 31.0 (10.3) & 1 (0.0521) & 0.220 \\
 303 & $\sim$ 1 (16.9) & 1.29 (21.8) & 0.01 (0.170) & 15.8 (134) & 1.53 (0.101) & 0.885 \\
 309 & $\sim$ 1 (1.50) & 1.40 (2.10) & 0.540 (0.813) & 43.1 (32.4) & 1.35 (0.0606) & 0.195 \\
 346 & $\sim$ 1 (0.724) & 2.05 (1.52) & 0.220 (0.188) & 66.1 (23.5) & $\sim 1$ (0.0809) & 0.322 \\
 434 & $\sim$ 1 (0.740) & 2.67 (2.04) & 0.592 (0.540) & 52.1 (19.3) & 1.23 (0.154) & 0.797 \\
 559 & $\sim$ 1 (24.2) & 1.31 (31.8) & 0.01 (0.242) & 34.3 (415) & 1.09 (0.198) & 1.03 \\
 597 & $\sim$ 1 (12.2) & 1.02 (12.4) & 0.01 (0.131) & 31.9 (194) & 1.17 (0.277) & 1.12 \\
 \textit{714*} & $\mathit{\sim}$ \textit{0} (\textit{60.4}) & \textit{0.798} ($\mathit{\sim 10^6}$) & \textit{0.830} ($\mathit{\sim 10^6}$) & \textit{364} ($\mathit{\sim 10^9}$) & \textit{2} ($\mathit{\sim 10^7}$) & \textit{1.97} \\
 768 & $\sim$ 1 (101) & 0.189 (19.2) & 0.01 (1.01) & 7.68 (389) & 1.63 (0.0600) & 0.980 \\
 839 & $\sim$ 1 (2.53) & 0.385 (0.974) & 0.393 (0.999) & 7.64 (9.67) & 2 (0.0729) & 1.02 \\
 \hline
\end{tabular}
\end{center}
\caption{Parameter values and corresponding $\chi_{\mathrm{min}}^2/N$ values of our fits in the hemisphere geometry for our sample of 17 SDSS-RM objects. The marginalized uncertainties of parameters are given in parentheses, and the cases where the model clearly fails are asterisked and italicized. For RM 17 specifically, we also quote the $\chi_{\mathrm{min}}^2/N$ where we have excluded data points from $1150 \ \rm{days} < t < 1250 \ \rm{days}$ in curly braces. The radius, $R_h$, is given in units of $R_s$, the Schwarzschild radius.}
\label{table:hemisphere_fit_param_vals}
\end{table*}

In addition, light curves of successfully-fit objects are poorly fit by reprocessing at certain times, once again suggesting that these reprocessing models alone are incomplete. For example, RM 17, Figure \ref{fig:case_study_lc}(a), has a sharp luminosity peak at $t \approx 1200$ days where the entire spectrum suddenly becomes much redder with the reddest bands rising to almost the same luminosity as the bluer bands \citep[see ][for more details on this object]{dexter2019}. As expected, neither reprocessing geometry is able to reproduce this significant increase in the redder bands as can be seen from Figure \ref{fig:rm17_fits}. Reprocessing alone is unable to explain such drastic changes in color, and the red light curve from our quantitative models significantly undershoots the observed data during this period. We quote the reduced minimum-$\chi^2$ values with and without (curly braces) data points from $1150 \ \rm{days} < t < 1250 \ \rm{days}$ in Tables \ref{table:hemisphere_fit_param_vals} and \ref{table:disk_fit_param_vals} emphasizing the significance of this deviation.

\subsection{Parameter Fit Results and Classification}

\begin{table*}[t!]
\begin{center}
\begin{tabular}{||c||c|c|c|c|c||c||} 
 \hline
 \multirow{2}{*}{RM-ID} & \multicolumn{6}{c||}{Thick Disk} \\
 {} & $H/R$ & $h_C$ $[R_g]$ & $\overline{L_C} \ [10^{45} \ \rm{erg} \ \rm{s}^{-1}]$ & $L_A/\overline{L_C}$ & $s$ & $\chi_{\mathrm{min}}^2/N$ \\
 \midrule
 12 & $\sim 0$ (0.773) & $\sim 20$ (19.3) & 1.46 (2.14) & 0.0710 (0.468) & $\sim 1$ (0.760) & 1.94 \\
 17 & 0.353 (0.304) & 2.94 (37.1) & $\sim 10$ (130) & 0.0257 (0.333) & 1.67 (0.0420) & 1.41 \{0.947\} \\
 \textit{32*} & \textit{0.122 (7.84)} & $\mathit{\sim 20}$ \textit{(4230)} & $\sim 10$ (1910) & 1.46 (279) & $\mathit{\sim 2}$ \textit{(1.65)} & \textit{3.51} \\
 \textit{105*} & $\mathit{\sim 0}$ \textit{(0.477)} & $\mathit{\sim 20}$ \textit{(6.76)} & \textit{0.962 (0.625)} & \textit{0.113 (0.144)} & $\mathit{\sim 1}$ \textit{(0.251)} & \textit{1.49} \\
 112 & 0.112 (0.305) & $\sim 20$ (3.59) & 1.53 (0.565) & 0.205 (0.108) & $\sim 1$ (0.119) & 1.31 \\
 143 & 0.325 (0.316) & $\sim 20$ (8.13) & 1.79 (0.921) & 0.01 (0.150) & 1.00 (0.247) & 0.820 \\
 160 & $\sim 0$ (1.00) & $\sim 20$ (18.5) & 0.257 (0.398) & 0.0196 (0.207) & $\sim 1$ (0.394) & 2.99 \\
 194 & $\sim 0$ (0.504) & $\sim 20$ (7.00) & 2.58 (1.80) & 0.136 (0.132) & $\sim 1$ (0.171) & 0.384 \\
 303 & 1.72 (1.27) & 6.08 (8.31) & 5.31 (15.2) & 0.01 (0.0251) & 1.58 (0.279) & 0.726 \\
 309 & 0.128 (0.467) & $\sim 20$ (12.2) & 2.41 (1.56) & 0.347 (0.252) & 1.55 (0.0463) & 0.250 \\
 346 & $\sim 0$ (0.689) & $\sim 20$ (16.3) & 4.44 (5.66) & 0.01 (0.236) & $\sim 1$ (0.627) & 0.953 \\
 434 & $\sim 0$ (0.419) & $\sim 20$ (11.2) & 8.28 (7.12) & 0.0101 (0.286) & $\sim 1$ (0.445) & 1.52 \\
 559 & 1.42 (1.07) & $\sim 20$ (67.9) & 1.41 (5.10) & 0.651 (2.78) & 1.82 (0.101) & 1.08 \\
 597 & 0.914 (1.20) & $\sim 20$ (33.0) & 1.26 (3.85) & 0.01 (1.80) & 1.19 (2.64) & 1.13 \\
 714 & 3.60 (4.43) & $\sim 20$ (19.7) & 0.791 (0.725) & 0.479 (3.27) & $\sim 1$ (0.759) & 0.667 \\
 768 & 4.37 (6.62) & 4.01 (1.64) & 3.25 (13.9) & 0.01 (0.0255) & 1.84 (0.0491) & 0.861 \\
 839 & 1.06 (0.572) & 0.588 (1.11) & $\sim 10$ (26.5) & 0.0224 (0.535) & $\sim 2$ (0.0649) & 0.955 \\
 \hline
\end{tabular}
\end{center}
\caption{Parameter values and corresponding $\chi_{\mathrm{min}}^2/N$ values of our fits in the thick-disk geometry for our sample of 17 SDSS-RM objects. The marginalized uncertainties of parameters are given in parentheses, and the cases where the model clearly fails are asterisked and italicized. For RM 17 specifically, we also quote the $\chi_{\mathrm{min}}^2/N$ where we have excluded data points from $1150 \ \rm{days} < t < 1250 \ \rm{days}$ in curly braces. The source height, $h_C$, is given in units of $R_g$, the gravitational radius.}
\label{table:disk_fit_param_vals}
\end{table*}

Tables \ref{table:hemisphere_fit_param_vals} and \ref{table:disk_fit_param_vals} present our parameter fit results for the hemisphere and disk geometries, respectively, for all 17 hypervariable quasars in our sample. Table \ref{table:summary_fit} shows the reduced minimum-$\chi^2$ values obtained in each geometry with the final two columns indicating the likely geometric classification --- whether the hemisphere model or thick-disk model is preferred, and the corresponding $\Delta$AIC value, indicating the statistical significance of the model preference. We note that many of the best-fit hemispheres have a much smaller radius than what can be reasonably expected for some reprocessing structure lying outside of the accretion disk. In these cases, the hemisphere corresponds to where the accretion disk would be emitting in the optical, suggesting that this reprocessing hemisphere is likely the accretion structure itself i.e. an extremely thick disk. In the case of RM 17 which is radio-loud, the hemisphere model could also be describing reprocessing by the base of the jet. We also note that the best-fit $H/R$ value for our disk geometry is likely underestimated due to our lack of treatment of multiple reprocessing in an elevated disk (see a more extensive discussion in Section \ref{sec:conclusion}).

\begin{center}
\begin{table*}
\begin{tabular}{||c||c|c||c|c||}
\hline
 RM-ID & Hemisphere $\chi^2_{\rm{min}}/N$ & Thick Disk $\chi^2_{\rm{min}}/N$ & Classification & $\Delta$AIC \\
 \hline
 12 & 0.907 & 1.94 & Hemisphere & 344 \\
 17 & 1.74 & 1.41 & Thick Disk & 187 \\
 32 & 0.284 & 3.51 & Hemisphere & 800 \\
 105 & 1.32 & 1.49 & - & - \\
 112 & 1.03 & 1.31 & Hemisphere & 113 \\
 143 & 0.805 & 0.820 & Hemisphere & 6.23 \\
 160 & 1.34 & 2.99 & Hemisphere & 838 \\
 194 & 0.220 & 0.384 & Hemisphere & 54.9 \\
 303 & 0.885 & 0.726 & Thick Disk & 66.7 \\
 309 & 0.195 & 0.250 & Hemisphere & 18.2 \\
 346 & 0.322 & 0.953 & Hemisphere & 210 \\
 434 & 0.797 & 1.52 & Hemisphere & 236 \\
 559 & 1.03 & 1.08 & Hemisphere & 18.0 \\
 597 & 1.12 & 1.13 & Hemisphere & 4.28 \\
 714 & 1.97 & 0.667 & Thick Disk & 657 \\
 768 & 0.980 & 0.861 & Thick Disk & 69.8 \\
 839 & 1.02 & 0.955 & Thick Disk & 34.0 \\
 \hline
\end{tabular}
\caption{$\chi_{\mathrm{min}}^2/N$ values of our minimum-$\chi^2$ fits in both the hemisphere and thick-disk geometries for our sample of 17 SDSS-RM objects. Each object's classification and the corresponding $\Delta$AIC, which indicates the statistical significance of the model preference, are listed in the final two columns.}
\label{table:summary_fit}
\end{table*}
\end{center}

\section{Conclusions} \label{sec:conclusion}

We examine the structures of the accretion in a sample of 17 hypervariable quasars from the SDSS catalog by considering thermal reprocessing models in alternate geometries with large covering areas: the hemisphere geometry and the thick-disk geometry. Our main results are summarized below:

\begin{itemize}
    \item All of the hypervariable quasars are best described by thick disks or hemispheres. Hypervariability likely requires a reprocessing structure that is vertically extended, unlike a standard \citet{shakura1973} thin disk.
    \item Of the 17 quasars, 11 are classified as a hemisphere, 5 are best-fit by a thick disk ($H/R > 0$), and one object (RM 105) is poorly fit by both quantitative models. Our classification scheme provides a first-order method of distinguishing between likely reprocessing geometries of hypervariable quasars, selecting an appropriate covering factor corresponding to their color evolution of variability.
    \item The failure of RM 105 and the sudden change in RM 17's color are both convincing examples where our simple reprocessing models cannot fully account for the observed hypervariability. This indicates incompleteness in the model.
\end{itemize}

\section{Discussion and Future Work} \label{sec:discussion}

While most objects in our sample were successfully modeled by either of the two geometries, there are certain important issues to be addressed before definitive conclusions can be drawn about the role of reprocessing in producing the optical spectra of quasars. While there are still many aspects of quasar accretion we do not understand, we hope that this study serves as a starting point for quasar classification and further examinations of the role of thermal reprocessing in quasars.

\subsection{Physical Picture of Our Reprocessing Geometries}

Our simplified geometric models are representative of plausible reprocessing structures surrounding quasars. The hemisphere model could correspond to a first-order approximation for out-of-plane material of significant density such that the optical depth is $\sim 1$, which can be caused by outflows/winds or warped/tilted disks. The thick-disk model could be attributed to enhanced thickness of disks elevated by magnetic fields, for instance. We have explored whether the optical spectrum of hypervariable quasars might be entirely dominated by thermal reprocessing.

\subsection{Data Analysis}

In our uncertainty analysis, we opt to estimate uncertainties empirically. This method incorporates both spectrophotometric errors and calibration uncertainties. We use the 2014 high-cadence data to estimate these empirical uncertainties, but the number of data points is too small ($N=32$) to estimate these uncertainties accurately. We fit a quadratic to model the intrinsic quasar variability in 2014, but this approach might not capture the variability entirely. Furthermore, we arbitrarily set a minimum fractional uncertainty of 10\%. A combination of these choices have led us to overestimate or underestimate quasar light curve uncertainties as indicated by the reduced minimum-$\chi^2$ values' deviation from unity. However, we note that our model selection relies only on comparing relative $\chi^2$, so our general results are unaffected by our choice of fractional uncertainty. As improvements are made in both data quantity and quality, we hope to improve our uncertainty characeterization as well. 

For the outlier-rejection algorithm, we assume that extreme quasar variability within 30 days can be excluded, but this choice is an arbitrary minimum timescale. In addition, this criterion is challenging to implement with a cadence of just 16 days (in the BOSS Spectrograph frame) in the years excluding 2014. We chose the wavelength range $4100 $--$ 9000$ \AA \ based on empirically estimated fractional uncertainties, which were significantly worse ($\sim 30 $--$ 50$\%) for wavelength bands $< 4100$ \AA \ and $> 9000$ \AA \ in the observed frame. We once again neglect spectrophotometric errors in this analysis.

Further improvements can be made to our process of estimating the driving light curve. \tanote{Firstly, we use linear interpolation for simplicity, but more sophisticated methods such as JAVELIN \citep{javelin} or CREAM \citep{cream} would be able to construct a more realistic driving light curve.} The ideal case is to obtain high cadence data in the X-ray or EUV to capture the driving light curve itself. This could be done through the STAR-X mission with its time domain X-ray/UV observations \citep{saha2017}, for instance. This program would better constrain our reprocessing models and significantly improve the classification scheme we introduced here, since the X-rays indeed appear to predict EUV well based on He II line widths \citep{timlin2021}. While we do not yet have this high cadence X-ray data for our sample, there are ways to better constrain our driving light curve in the immediate future. In particular, the mean luminosity, $\overline{L_C}$, and the stretch factor, $s$, can be better constrained by performing one-time X-ray luminosity measurements through XMM-Newton or Chandra. The uncertainty in the time-delay, $t_0$, can be reduced by geometry from estimating the location of the bluest light curve emission relative to the central source, for instance through recent disk-size measurements \citep[e.g.,][]{homayouni2019}.

\subsection{Host-Galaxy Contributions and Reddening}
\label{sec:host_galaxy_reddening}

There are significant uncertainties introduced by the host-galaxy subtraction. \citet{shappee2014} performs model fits with an additional constant luminosity term in each wavelength band which allows for many additional degrees of freedom, whereas we initially subtract an expected SED \citep{shen2015}. Although we include the subtracted host-galaxy light when characterizing the uncertainty of data points for our minimum-$\chi^2$ analysis, inaccurate host-galaxy subtraction could alter our light curves' fractional variability which in turn affects our model fits.

Additionally, there are sources of reddening that are unaccounted for. While Galactic extinction is negligible in the SDSS field, reddening due to dust in the quasar's host-galaxy and intergalactic reddening are ignored. While constant reddening should not affect the variability amplitudes, it does affect the relative luminosities between wavelength bands and consequently our model fit results. In addition, a hypervariable quasar with dramatically variable reddening has recently been discovered from SDSS-V \citep{zeltyn2022}. Variable reddening would alter the variability amplitude and the likely reprocessing geometry of the hypervariable quasar. Our preliminary results show that extremely large reddening of $E(B-V) \sim 0.1 $--$ 0.3$ mag is required to make significant changes to the model fit results, but smaller $E(B-V)$ could still alter our classification results for certain objects.

\subsection{Quantitative Models and Classification}

We only consider singly-reprocessed light for the thick-disk case, but in reality, light from the central source can be reprocessed multiple times by our geometry before reaching the observer. Since multiple reprocessing becomes increasingly more likely as the disk becomes more elevated, this approach effectively penalizes highly-elevated disks as they become less efficient. This limitation explains our classification results where some objects prefer a model with $H/R \sim 0$ but with $L_A \ll L_C$ which is physically unreasonable. These objects may very well be successfully modeled by an elevated disk with multiple reprocessing accounted for. Multiple reprocessing, however, is challenging to model, since the light travel time-delay becomes increasingly complex. Furthermore, the standard thin disk \citep{shakura1973} was used to model the accretion luminosity, but a slim disk \citep{abramowicz1988} might be a better approximation.

The current hemisphere model is inconsistent with the observed mean AGN spectrum, since we only consider a reprocessor at a single radius. A more realistic reprocessor could be multi-temperature and/or radially extended. The hemisphere geometry is most consistent with a physical picture where there is some reprocessing out-of-plane material lying outside of much of the accretion structure. However, as Table \ref{table:hemisphere_fit_param_vals} reveals, many of the reprocessing hemispheres selected are quite small with radii corresponding to where the accretion disk would be emitting in the optical, according to standard thin disk theory \citep{shakura1973}. This result means that the reprocessing hemisphere likely represents the accretion structure itself. In future iterations of this analysis where multiple reprocessing in a thick disk is fully taken into account, the hemisphere geometry may be disfavored. \tanote{For simplicity, we assumed a face-on geometry for both the hemisphere and thick-disk models, but this assumption can be relaxed in future efforts.} Additionally, we will make quantitative comparisons between our fit results and SDSS-RM optical disk-size measurements \citep[e.g.,][]{homayouni2019} to better understand the likely reprocessing geometry of our targets.

\section{Acknowledgements}

\tanote{We thank the anonymous referee for their suggestions which greatly improved the quality of our paper.}

TA and JD were supported in part by NSF grant AST-1909711, and by an Alfred P. Sloan Research Fellowship (JD). WNB was supported by NSF grant AST-2106990. LCH was supported by the National Science Foundation of China (11721303, 11991052, 12011540375) and the China Manned Space Project (CMS-CSST-2021-A04, CMS-CSST-2021-A06). YS acknowledges support from NSF grant AST-2009947. JRT acknowledges support from NSF grants CAREER-1945546, AST-2009539, and AST-2108668. 

Funding for the Sloan Digital Sky Survey IV has been provided by the Alfred P. Sloan Foundation, the U.S. Department of Energy Office of Science, and the Participating Institutions. SDSS-IV acknowledges
support and resources from the Center for High-Performance Computing at the University of Utah. The SDSS web site is www.sdss.org.

SDSS-IV is managed by the Astrophysical Research Consortium for the 
Participating Institutions of the SDSS Collaboration including the 
Brazilian Participation Group, the Carnegie Institution for Science, 
Carnegie Mellon University, the Chilean Participation Group, the French Participation Group, Harvard-Smithsonian Center for Astrophysics, 
Instituto de Astrof\'isica de Canarias, The Johns Hopkins University, Kavli Institute for the Physics and Mathematics of the Universe (IPMU) / 
University of Tokyo, the Korean Participation Group, Lawrence Berkeley National Laboratory, 
Leibniz Institut f\"ur Astrophysik Potsdam (AIP),  
Max-Planck-Institut f\"ur Astronomie (MPIA Heidelberg), 
Max-Planck-Institut f\"ur Astrophysik (MPA Garching), 
Max-Planck-Institut f\"ur Extraterrestrische Physik (MPE), 
National Astronomical Observatories of China, New Mexico State University, 
New York University, University of Notre Dame, 
Observat\'ario Nacional / MCTI, The Ohio State University, 
Pennsylvania State University, Shanghai Astronomical Observatory, 
United Kingdom Participation Group,
Universidad Nacional Aut\'onoma de M\'exico, University of Arizona, 
University of Colorado Boulder, University of Oxford, University of Portsmouth, 
University of Utah, University of Virginia, University of Washington, University of Wisconsin, 
Vanderbilt University, and Yale University.

\bibliographystyle{aasjournal}
\bibliography{main}

\end{document}